\pdfoutput=1
\documentclass[fleqn, usenatbib, usedcolumn]{mnras}

\newif\iflocal
\localfalse

\usepackage{graphicx}
\usepackage{xspace}
\usepackage{amsmath}
\usepackage{framed} 
\usepackage{txfonts}
\usepackage{epstopdf}
\usepackage{color}
\usepackage{rotating}
\usepackage{natbib}
\usepackage{ulem}
\usepackage{xspace}
\usepackage{sistyle}

\iflocal
\def\includedir{/Users/benedito/University/docs/latex}
\def\figdir{figs}
\else
\def\includedir{.}
\def\figdir{.}
\fi

\newcommand{\figmn}[1]{Fig.~\ref{#1}\xspace}
\newcommand{\eqmn}[1]{equation~(\ref{#1})\xspace}
\newcommand{\eqmnb}[1]{equation~\ref{#1}\xspace}

\def\gtsima{$\; \buildrel > \over \sim \;$}
\def\ltsima{$\; \buildrel < \over \sim \;$}
\def\prosima{$\; \buildrel \propto \over \sim \;$}
\def\gsim{\lower.7ex\hbox{\gtsima}}
\def\lsim{\lower.7ex\hbox{\ltsima}}
\def\simgt{\lower.7ex\hbox{\gtsima}}
\def\simlt{\lower.7ex\hbox{\ltsima}}
\def\simpr{\lower.7ex\hbox{\prosima}}

\newcommand{\dnorm}[2]{\frac{{\rm d} #1}{{\rm d} #2}}
\newcommand{\dnorminl}[2]{{\rm d} #1 / {\rm d} #2}
\newcommand{\dnormtwo}[2]{\frac{{\rm d}^2 #1}{{\rm d} #2^2}}
\newcommand{\dnorminltwo}[2]{{\rm d}^2 #1 / {\rm d} #2^2}

\newcommand{\ndof}{N_{\rm dof}}
\newcommand\chidof{\chi^2/\ndof}

\newcommand{\gyr}{{\rm Gyr}}

\newcommand{\msun}{{M_{\odot}}}
\newcommand{\msunh}{h^{-1} M_\odot}

\newcommand{\LCDM}{$\Lambda$CDM\xspace}

\newcommand{\rhom}{\rho_{\rm m}}

\def\sparta{\textsc{Sparta}\xspace}

\def\colossus{\textsc{Colossus}\xspace}

\def\planck{Planck\xspace}
\def\wmap{WMAP7\xspace}

\def\erebos{Erebos\xspace}

\def\rmd{{\rm d}}

\def\rmi{{\rm i}}

\def\rms{{\rm s}}
\def\rmt{{\rm t}}

\def\deltac{\delta_{\rm c}}
\def\rs{r_{\rm s}}

\def\rhoorb{\rho_{\rm orb}}

\def\rvir{R_{\rm vir}}

\def\mtom{M_{\rm 200m}}
\def\rtom{R_{\rm 200m}}
\def\ctom{c_{\rm 200m}}

\def\ntom{N_{\rm 200m}}

\def\rtoc{R_{\rm 200c}}

\def\mtombnd{M_{\rm 200m,bnd}}
\def\mtomall{M_{\rm 200m,all}}

\newcommand{\rhos}{\rho_\rms}
\newcommand{\rt}{r_\rmt}

\newcommand{\rrs}{\left( \frac{r}{\rs} \right)}
\newcommand{\rrt}{\left( \frac{r}{\rt} \right)}
\newcommand{\rsrt}{\left( \frac{\rs}{\rt} \right)}

\newcommand{\rrtom}{\left( \frac{r}{\rtom} \right)}
\newcommand{\rR}{\left( \frac{r}{R} \right)}

\newcommand{\rrsa}{\rrs^\alpha}
\newcommand{\rrse}{\rrs^\eta}
\newcommand{\rrtb}{\rrt^\beta}

\newcommand{\rsrtb}{\rsrt^\beta}

\newcommand{\delone}{\delta_{\rm 1}}
\newcommand{\delmax}{\delta_{\rm max}}

\usepackage{etoolbox}
\makeatletter
\patchcmd{\NAT@citex}
  {\@citea\NAT@hyper@{\NAT@nmfmt{\NAT@nm}\NAT@date}}
  {\@citea\NAT@nmfmt{\NAT@nm}\NAT@hyper@{\NAT@date}}
  {}
  {}
\patchcmd{\NAT@citex}
  {\@citea\NAT@hyper@{%
     \NAT@nmfmt{\NAT@nm}%
     \hyper@natlinkbreak{\NAT@aysep\NAT@spacechar}{\@citeb\@extra@b@citeb}%
     \NAT@date}}
  {\@citea\NAT@nmfmt{\NAT@nm}%
   \NAT@aysep\NAT@spacechar%
   \NAT@hyper@{\NAT@date}}
  {}
  {}
\patchcmd{\NAT@citex}
  {\@citea\NAT@hyper@{%
     \NAT@nmfmt{\NAT@nm}%
     \hyper@natlinkbreak{\NAT@spacechar\NAT@@open\if*#1*\else#1\NAT@spacechar\fi}%
       {\@citeb\@extra@b@citeb}%
     \NAT@date}}
  {\@citea\NAT@nmfmt{\NAT@nm}%
   \NAT@spacechar\NAT@@open\if*#1*\else#1\NAT@spacechar\fi%
   \NAT@hyper@{\NAT@date}}
  {}
  {}
\makeatother

\iflocal
\def\figdir{figs}
\else
\def\figdir{.}
\fi

\defcitealias{diemer_13_scalingrel}{DKM13}
\defcitealias{diemer_14}{DK14}
\defcitealias{diemer_15}{DK15}
\defcitealias{diemer_22_prof1}{I}

\newcommand{\paperone}{Paper \citetalias{diemer_22_prof1}\xspace}
\newcommand{\paperthree}{Paper III\xspace}
\newcommand{\dkft}{\citetalias{diemer_14}\xspace}

\title[Dynamics-based halo density profiles -- II.]{A dynamics-based density profile for dark haloes -- II. Fitting function}

\author[Diemer]{Benedikt Diemer\thanks{Email: \href{mailto:diemer@umd.edu}{diemer@umd.edu}}
\vspace{1mm}
\\
Department of Astronomy, University of Maryland, College Park, MD 20742, USA \\
}

\date{Accepted 2022 December 18. Received 2022 November 15; in original form 2022 May 6}
\pubyear{2023}

\begin{document}
\label{firstpage}
\pagerange{\pageref{firstpage}--\pageref{lastpage}}
\maketitle


\begin{abstract}
The density profiles of dark matter haloes are commonly described by fitting functions such as the NFW or Einasto models, but these approximations break down in the transition region where halos become dominated by newly accreting matter. Here we present a simple, accurate new fitting function that is inspired by the asymptotic shapes of the separate orbiting and infalling halo components. The orbiting term is described as a truncated Einasto profile, $\rhoorb \propto \exp \left[-2/\alpha\ (r / \rs)^\alpha - 1/\beta\ (r / \rt)^\beta \right]$, with a five-parameter space of normalization, physically distinct scale and truncation radii, and $\alpha$ and $\beta$, which control how rapidly the profiles steepen. The infalling profile is modelled as a power law in overdensity that smoothly transitions to a constant at the halo centre. We show that these formulae fit the averaged, total profiles in simulations to about 5\% accuracy across almost all of an expansive parameter space in halo mass, redshift, cosmology, and accretion rate. When fixing $\alpha = 0.18$ and $\beta = 3$, the formula becomes a three-parameter model that fits individual halos better than the Einasto profile on average. By analogy with King profiles, we show that the sharp truncation resembles a cut-off in binding energy.
\end{abstract}



\section{Introduction}
\label{sec:intro}

The shapes that make up the cosmic web of dark matter are generally too complex to be described mathematically \citep[e.g.][]{doroshkevich_78, bond_96_filaments}. However, the densest structures, or haloes, collapse into relatively uniform and roughly spherical shapes. Describing haloes analytically is critical because they contain about half the dark and most of the observable baryons in the Universe. The most common description are spherically averaged density profiles, which conveniently summarize simulation results \citep{dubinski_91}, serve as the basis of galaxy formation models \citep{wechsler_18}, underlie descriptions of large-scale structure via the halo model \citep{cooray_02}, are used to fit observed profile data \citep[e.g.,][]{courteau_14}, and provide a parameter space in which to compare such observations to theory and simulations \citep[e.g.,][]{umetsu_20_review, eckert_22_einasto}. 

To facilitate these applications, numerous fitting functions have been proposed, although many were originally intended for galaxies rather than haloes. The three-parameter \citet{einasto_65, einasto_69} profile features a slope that smoothly steepens with radius and thus contains a finite mass. Many later proposals are essentially double power laws that smoothly transition between two slopes around a scale radius, $\rs$ \citep{dehnen_93}. This group includes the profiles of \citealt{jaffe_83} (with inner slope $-2$ and outer slope $-4$), \citealt{hernquist_90} ($-1$ and $-4$), \citealt{burkert_95} ($0$ and $-3$), \citealt{navarro_95, navarro_96, navarro_97} (hereafter NFW, $-1$ and $-3$), and \citealt{moore_99_collapse} ( $-1.5$ and $-3$, see also \citealt{ghigna_00}). Numerous studies have found the extra parameter of the Einasto profile to be justified by a superior fit to simulation results \citep{navarro_04, navarro_10, graham_06, merritt_06, gao_08, stadel_09, ludlow_11, wang_20_zoom}. Naturally, the fit of power-law models can be improved by adding more slope parameters \citep{zhao_96, dekel_17, freundlich_20}. Finally, an entire different class of models can be derived by assuming a distribution of particle energies and computing the corresponding density structure \citep{king_66, shapiro_99, hjorth_10_darkexp1, pontzen_13}.

All of these models were intended to fit density profiles out to roughly the virial radius, a focus that makes sense given that galaxy formation happens at much smaller radii. For example, one of the key debates has been whether the central slope of the profiles is characteristic of a flat core or a power-law cusp \citep[e.g.,][]{deblok_10, teyssier_13, dicintio_14, genina_18}. A second reason is that the nature of the profiles changes fundamentally beyond the virial radius, where the profile becomes dominated by particles that are falling into the halo for the first time. In this transition region, the profile shape is the result of a complex interplay between the orbiting and infalling terms, an inherently dynamical distinction that cannot easily be made based on the profiles alone (\citealt{fukushige_01}, \citealt{diemand_08}). As a result, conventional fitting functions trained on the orbiting (or one-halo) term struggle to fit simulated or observed profiles in this regime \citep{becker_11, oguri_11}. While some models for the outer profiles have been proposed \citep{prada_06, betancortrijo_06, tavio_08, baltz_09}, they fail to capture the detailed profile shapes \citep[][hereafter \dkft]{diemer_14}.

Despite these complications, the transition region and the outer profiles have recently gained renewed relevance because they are much less affected by baryons and because they are sensitive to otherwise inaccessible halo properties. For example, the mass accretion rate sets the position of the edge of the orbiting term, which is also known as the splashback radius \citep[\dkft,][]{adhikari_14, more_15}. Moreover, the outer profiles have become observationally accessible via the density of satellite galaxies and the weak lensing signal around clusters \citep[e.g.,][]{more_16, baxter_17, chang_18, shin_19_rsp, murata_20, bianconi_21}. For their inferences, all of these works have relied on the fitting function of \dkft, an Einasto profile multiplied by a power-law cut-off term with variable truncation radius, sharpness, and asymptotic slope (a total of $6$ free parameters). This flexible function fits averaged profiles to 5--10\% accuracy including the transition region, but its design suffered from the same fundamental issue as previous models: at its truncation, the orbiting term is concealed by the infalling term, and its asymptotic shape was thus unknown. As a result, the \dkft fitting function is hard to interpret physically. First, the large number of parameters causes well-known degeneracies that necessitate informative priors \citep[\dkft;][]{baxter_17, chang_18, umetsu_17}. Second, the steepening term approaches a somewhat arbitrary slope that does not actually reflect the shape of the orbiting term. Some observational works have extrapolated this shape based on \dkft fits \citep[e.g.][]{baxter_17, shin_21}, but it is not clear that the results are physically meaningful. Third, the slope of the \dkft profile is a complex function, which makes it difficult to establish the relationship between the profile parameters and physical features such as the splashback radius.

In this paper, we set out to design a more physically motivated fitting function. For the first time, we can draw inspiration from profiles that have been dynamically split into orbiting and infalling particles \citep[using the algorithm presented in][hereafter \paperone; see also \citealt{garcia_22}]{diemer_22_prof1}. Similar splits have recently been explored in observational data using galaxy properties such as colour \citep{baxter_17, adhikari_21, shin_21, odonnell_22, aung_22, dacunha_22, oneil_22}, necessitating a fitting function that captures the shapes of the orbiting and infalling terms. We require that this fitting function should 1) describe the profiles accurately even in the transition region, 2) be flexible enough to apply to individual halos and to stacks with different selection criteria, 3) work across all halo masses, redshifts, cosmologies, and accretion rates, 4) rely on as few free parameters as possible with clear, physical interpretations, and 5) exhibit minimal parameter degeneracies. We achieve these goals with a $5$-parameter Einasto-like form with a truncated exponent. Our aim is not necessarily to achieve more accurate fits than \dkft because there are complex variations in the averaged profiles that make it difficult to systematically improve on a 5--10\% fit. Moreover, baryons affect the profiles at roughly this level even at large radii \citep[e.g.,][]{velliscig_14, schneider_19}. Thus, we instead focus on creating a parameter space that can be used to meaningfully describe observed and simulated profiles. In this second paper of the series, we present and test the fitting function. In Diemer (in preparation; hereafter \paperthree), we analyse the resulting parameter space and connect it to halo physics such as accretion histories.

The paper is structured as follows. In Section~\ref{sec:models}, we mathematically describe previous fitting functions and the new models. We summarize our numerical methods in Section~\ref{sec:methods}, largely referring the reader to \paperone. We assess the quality of our fits to stacked and individual halo profiles in Section~\ref{sec:results}, and we compare to previously proposed models in Section~\ref{sec:comp}. Section~\ref{sec:conclusion} summarizes our results. We discuss an alternative model variant in Appendix~\ref{sec:app:modelb} and derive additional mathematical properties in Appendices~\ref{sec:app:derivs} and \ref{sec:app:comp_df}. Supplementary figures are provided online on the author's website at \href{http://www.benediktdiemer.com/data/}{benediktdiemer.com/data}. Our fitting function is implemented in the publicly available code \colossus \citep{diemer_18_colossus}.

Throughout the paper, we follow the notation of \paperone. We normalize halo radii by $\rtom$, which encloses an average of $200$ times the mean density of the Universe, $\rhom$, including all particles (bound or unbound). We denote the mass enclosed within this radius as $\mtom$. We mostly express mass as peak height, $\nu \equiv \deltac / \sigma(\mtom)$, where $\deltac = 1.686$ is the critical density for collapse \citep{gunn_72} and $\sigma(\mtom)$ is the variance of the linear power spectrum on the Lagrangian scale corresponding to $\mtom$. We define the logarithmic mass accretion rate $\Gamma \equiv \Delta \ln \mtom / \Delta \ln a$, measured over one dynamical time (which we define as the mass-independent halo crossing time, about $5\ \gyr$ at $z = 0$). More details on these calculations are given in \paperone.


\section{Profile models}
\label{sec:models}

In this section, we introduce new fitting functions for the orbiting and infalling components of density profiles, which can be combined to fit the entire profile. We describe the general logic of constructing models for the orbiting term in Section~\ref{sec:models:orb}, previously proposed models in Sections~\ref{sec:models:einasto} and \ref{sec:models:dk14}, and our new model in Section~\ref{sec:models_orb:moda}. We introduce a slightly different version of the same model in Appendix~\ref{sec:app:modelb}. We also discuss previous and new models for the infalling profile in Sections~\ref{sec:models:inf} and \ref{sec:models:plmk}. The derivatives of the models with respect to their parameters are given in Appendix~\ref{sec:app:derivs}.

\subsection{Generalized exponential profiles}
\label{sec:models:orb}

Our new function for the orbiting term can be thought of as a generalization of the Einasto profile, where the slope smoothly changes with radius. We write such profiles as the exponential of a radial function $S(r)$,
\begin{equation}
\label{eq:models:einasto}
\rho_{\rm orb} = \rhos e^{S(r)} \,.
\end{equation}
This family of profiles is more intuitively understood by considering the logarithmic slope $\gamma(r)$,
\begin{equation}
\gamma \equiv \dnorm{\ln \rho}{\ln r} = r \dnorm{S}{r} \,.
\end{equation}
This expression demonstrates the power of \eqmn{eq:models:einasto}: we can construct a density profile by defining its logarithmic slope and integrating to get $S(r)$,
\begin{equation}
S(r) = \int \frac{\gamma(r)}{r} \rmd r \,.
\end{equation}
Clearly, one condition for the fitting function to be useful is that $\gamma/r$ must integrate to a reasonably simple, analytical expression. For example, exponential cut-offs in the slope such as $\gamma \propto \exp(r / \rt)$ are problematic because their integrals (if solvable) involve Gamma functions or other complex mathematical expressions.

\subsection{The Einasto profile}
\label{sec:models:einasto}

\begin{figure*}
\centering
\includegraphics[trim =  2mm 9mm 0mm 3mm, clip, width=\textwidth]{\figdir/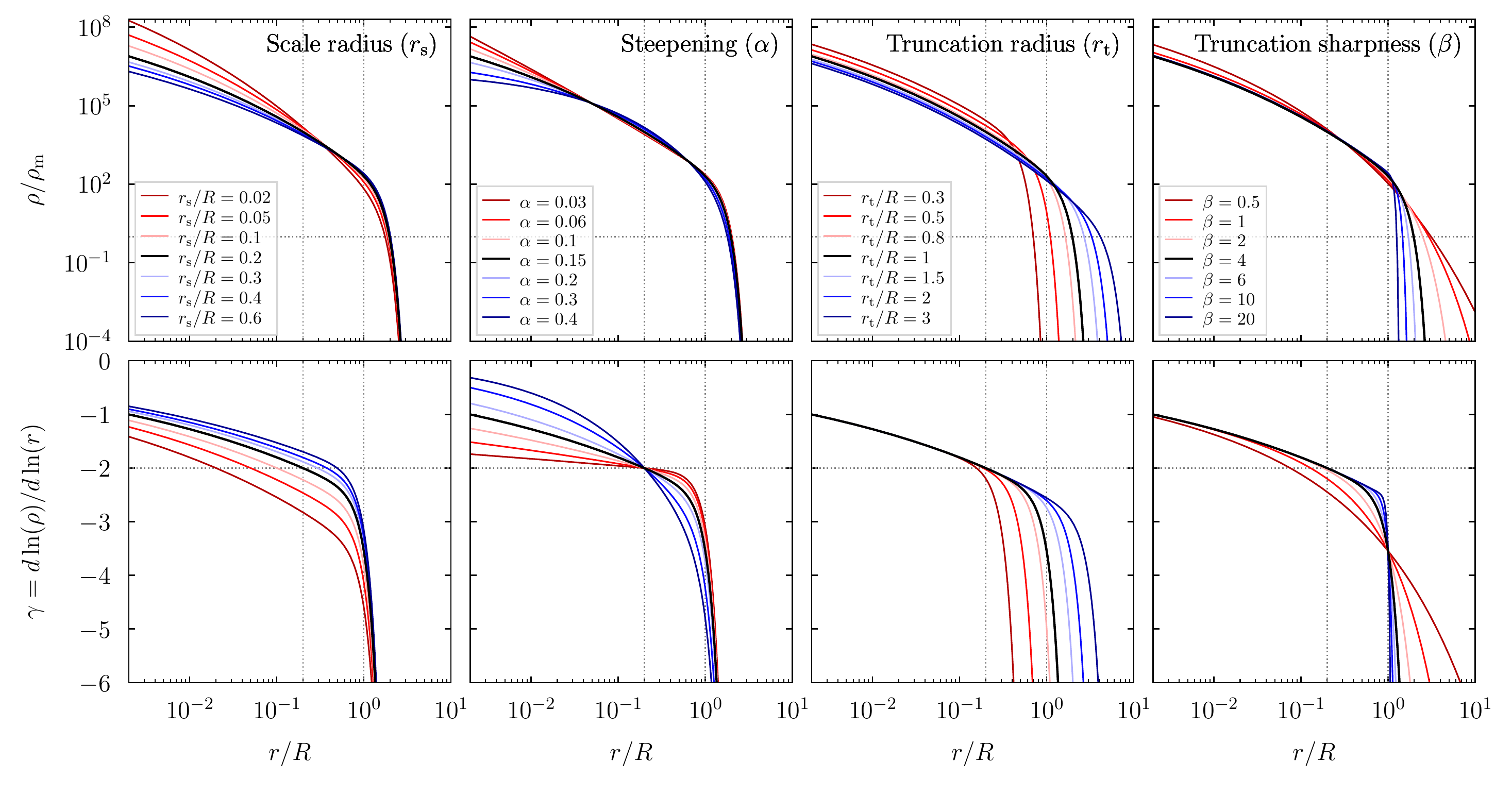}
\caption{Effect of the free parameters in our model for the orbiting density profile (top) and its logarithmic slope (bottom). Each column shows variations of a single parameter; $\rhos$ is omitted as it corresponds to a simple normalization. All profiles are normalized by some radius $R$ and integrate to the same $M(R)$. The black lines show the same fiducial profile in each column ($\rs = 0.2\ R$, $\alpha = 0.15$, $\rt = R$, $\beta = 4$), and the dotted vertical lines highlight the fiducial $\rs$ and $\rt$. The horizontal dashed lines mark the mean density in the top panels and a slope of $-2$ in the bottom panels. All slopes converge to $\gamma(0) = 0$ at the halo centre, but we would need to extend the plot to much smaller radii to see this effect.}
\label{fig:moda_pars}
\end{figure*}

We begin by reviewing the Einasto profile in light of the discussion above. Its defining feature is a slope of $\gamma = -2 (r / \rs)^\alpha$.
The parameter $\alpha$ determines how rapidly the profile steepens, and the meaning of the scale radius is that $\gamma(\rs) = -2$. Moreover, the Einasto profile leads to a core at small radii, $\gamma(0) = 0$. The slope integrates to $S(r) = -(2 / \alpha) (r / \rs)^\alpha + C$
We can choose to set $C = 0$, in which case $\rhos$ describes the density at $r = 0$, but the profile is more commonly written with $C = 2 / \alpha$,
\begin{equation}
\label{eq:models:einasto:s}
S(r) = -\frac{2}{\alpha} \left[ \rrsa - 1 \right] \,,
\end{equation}
such that $S(\rs) = 0$ and thus $\rho(\rs) = \rhos$. In practice, we find that the $C \neq 0$ version of the Einasto profile behaves better in least-squares fits because it does not allow a degeneracy between $\rhos$, $\rs$, and $\alpha$ (although the latter two can still be degenerate at fixed $\rhos$).

\subsection{The DK14 profile}
\label{sec:models:dk14}

We briefly review the \dkft model because we will later use its performance as a benchmark for our new model. The model is
\begin{equation}
\label{eq:models:dk14}
\rho_{\rm DK14} = \rho_{\rm Einasto} \left[1 + \left( \frac{r}{\rt} \right)^\beta \right]^{-\frac{\gamma}{\beta}} \,,
\end{equation}
where the steepening term suppresses the density at $r \gsim \rt$ as a power law with slope $-\gamma$, while $\beta$ governs the sharpness of the truncation. Combined with the Einasto parameters, this function has six free parameters, but \dkft suggested fixing $(\beta, \gamma)$ to $(4, 8)$ in fits to mass-selected samples and to $(6, 4)$ when the sample is also selected by mass accretion rate (they did not investigate fits to individual halos). The main disadvantage of \eqmn{eq:models:dk14} is that the $\beta$ and $\gamma$ parameters do not have a clear physical meaning because the asymptotic slope of the overall orbiting profile depends on $\rs$, $\rt$, $\alpha$, and $\gamma$ in a complex fashion. Setting $\beta$ and $\gamma$ to fixed values reduces the profile to a four-parameter fit with a meaningful truncation radius parameter $\rt$, but it no longer allows for a varying shape of the truncation term. 

\subsection{The new model: truncated exponential}
\label{sec:models_orb:moda}

\begin{figure*}
\centering
\includegraphics[trim =  2mm 9mm 0mm 3mm, clip, width=\textwidth]{\figdir/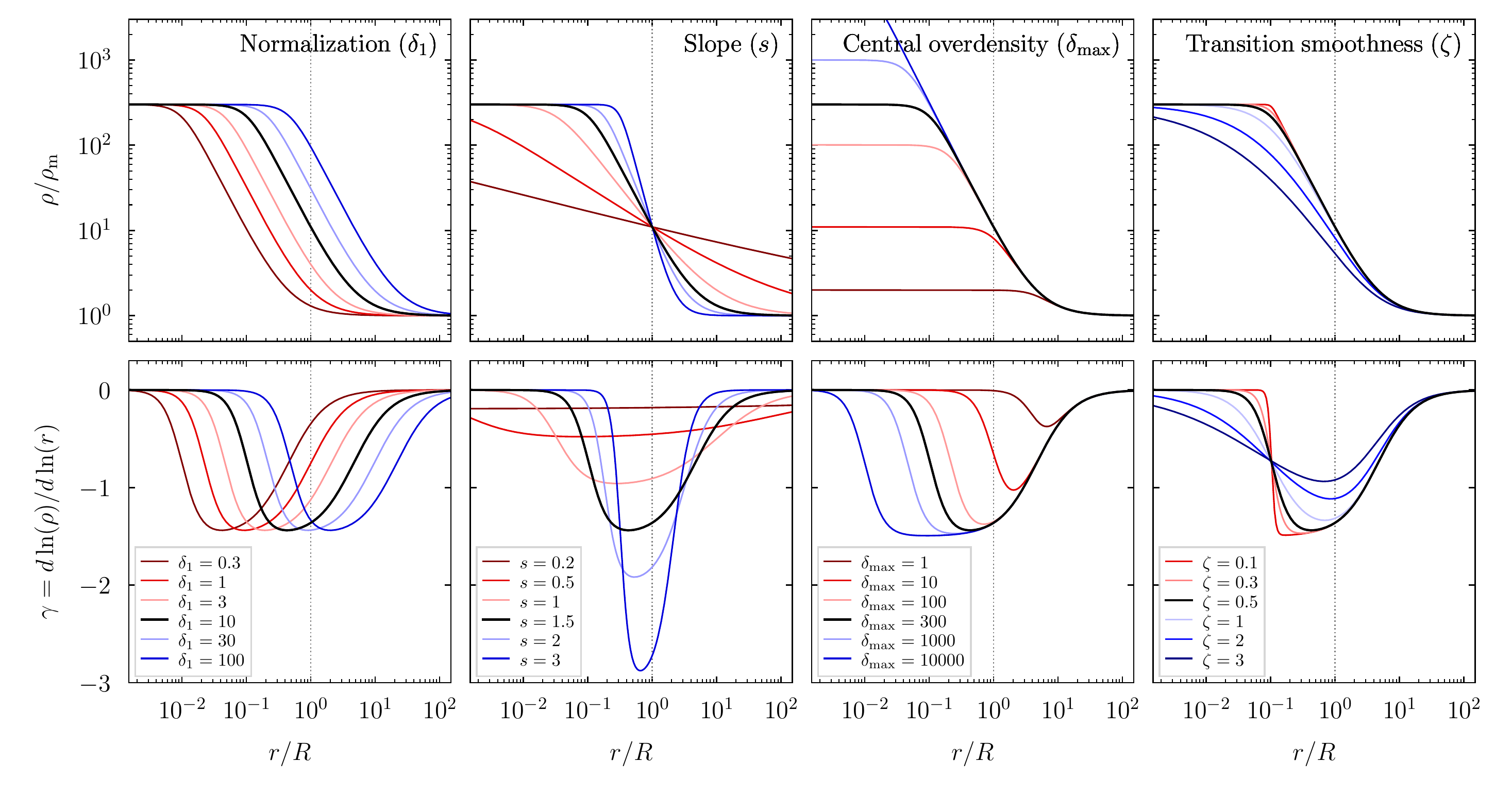}
\caption{Same as Fig.~\ref{fig:moda_pars} but for the new infalling profile model, which smoothly transitions from a power law to an asymptotic central density. The parameters are the normalization at radius $R$, the slope of the power law, the maximum central overdensity, and the transition smoothness. The $\delmax$ parameter can affect the overdensity at $R$ if it is sufficiently small.}
\label{fig:plmk_pars}
\end{figure*}

In \paperone, we confirmed that the orbiting profile has two characteristic radial scales, which can be captured by the scale and the truncation radii. Moreover, we found that the sharpness of the truncation varies with halo properties and that the profiles decline with an exponential-like, accelerating slope (rather than a constant, power-law slope). Inspired by these observations, we construct a model by adding a truncation term into the slope, 
\begin{equation}
\gamma(r) = -2 \rrsa - \rrtb \,.
\end{equation}
While similar in spirit, this model significantly differs from the `truncated Sersic' and `broken exponential' models that are popular in the observational literature \citep[e.g.][]{peng_10_images, erwin_15}. We find the corresponding profile by integrating $\gamma / r$,
\begin{equation}
\label{eq:models:moda:s}
S(r) = -\frac{2}{\alpha} \rrsa -\frac{1}{\beta} \rrtb + C \,.
\end{equation}
For simplicity, we could set $C = 0$, in which case $\rhos$ would become the density at the centre. A more elegant form is obtained by setting the integration constant such that $\rho(\rs) = \rhos$,
\begin{equation}
\label{eq:models:moda:c}
C = \frac{2}{\alpha} + \frac{1}{\beta} \rsrtb \,,
\end{equation}
and thus
\begin{equation}
\label{eq:models:moda:s2}
S(r) = -\frac{2}{\alpha} \left[ \rrsa - 1 \right] -\frac{1}{\beta} \left[ \rrtb - \rsrtb \right] \,.
\end{equation}
We use this parametrisation throughout because the density at $\rs$ is numerically constrained, whereas the central density represents an extrapolation. However, we give derivatives with respect to the free parameters for both variants in Appendix~\ref{sec:app:derivs:moda}. In \figmn{fig:moda_pars}, we visually explore how the free parameters affect the shape of our model. Throughout the rest of the paper, we show that \eqmn{eq:models:moda:s2} describes the orbiting profiles with great accuracy. 

The model does, however, have two minor shortcomings. First, the truncation term `breaks' the meaning of the scale radius because the slope at $\rs$ is now $\gamma(\rs) = -2 - (\rs / \rt)^\beta$ instead of $-2$. The difference is small in most cases since $\rt > \rs$ and generally $\beta \gg 1$, but it can manifest itself for extreme parameter values (red lines in the right two columns of \figmn{fig:moda_pars}). We note that the $\gamma(\rs) = -2$ condition will likely be violated regardless as soon as an infalling profile is added. Nonetheless, in Appendix~\ref{sec:app:modelb} we present a model variant that enforces $\gamma(\rs) = -2$ at the expense of an additional term in $S(r)$. Both models give the same fit and best-fit parameters for virtually all profiles, but the variant can be preferable in cases where $\rs$ and $\alpha$ are strongly degenerate. For either model, the scale radii (and thus concentrations) are directly comparable to those from Einasto fits, but we expect small, systematic shifts similar in scale to the differences between NFW and Einasto concentrations \citep{dutton_14} or differences due to the fitting procedure \citep{dooley_14}.

The second shortcoming is that the integrals of $\rho$, namely the enclosed mass $M(r)$, the projected density $\Sigma(R)$, and the lensing signal $\Delta \Sigma$, cannot be computed analytically. Similar issues hamper even the simpler Einasto profile, where $M(r)$ is a relatively complicated analytical expression \citep{cardone_05, retanamontenegro_12} and $\Sigma(R)$ can only be approximated \citep{dhar_10, dhar_21}. We will investigate similar approximate solutions for the new model in future work, and we have provided fast, numerical solution in the \colossus code \citep{diemer_18_colossus}.

\subsection{The infalling profile: previous models}
\label{sec:models:inf}

Before shells of dark matter begin to cross, the overdensity due to non-linearly infalling matter is expected to roughly follow a power law with a slope of $-3/2$ \citep{bertschinger_85}. At large radii, the statistical contribution from large-scale structure comes to dominate, but this transition happens at radii larger than the $10\ \rtom$ we consider (\paperone). Thus, it is not surprising that \dkft found the profiles to be well-described by a simple power law in overdensity, 
\begin{equation}
\label{eq:models:pl}
\rho(r) = \rhom(z) \left[ \delone \rR^{-s} + 1 \right] \,,
\end{equation}
where $\delone$ represents the normalisation at radius $R$ and $-s$ the slope. We will set $R = \rtom$ throughout, but any other pivot radius could be chosen instead. One issue with \eqmn{eq:models:pl} is that it can reach arbitrarily high values at small $r$, which is clearly unphysical.  We can prevent this artefact by introducing a maximum overdensity \citep[e.g.,][]{diemer_18_colossus},
\begin{equation}
\label{eq:models:plm0}
\rho(r) = \rhom \left( \delone \left[ \frac{\delone}{\delmax} + \rR^{s} \right]^{-1} + 1 \right) \,.
\end{equation}
This expression approaches $\rho \rightarrow \rhom (\delmax + 1)$ as $r \rightarrow 0$. The exact value of $\delmax$ does not matter because it is reached at radii where the infalling profile dominates by orders of magnitude. 

\subsection{New infalling model: power law with smooth transition}
\label{sec:models:plmk}

Inspecting the infalling profiles in \paperone, we found that they do indeed seem to approach a constant density at small radii, but the transition to this value happens less sharply than suggested by \eqmn{eq:models:plm0}. We thus introduce a transition smoothness parameter, $\zeta$,
\begin{equation}
\label{eq:models:plmk}
\rho(r) = \rhom \left( \delone \left[ \left( \frac{\delone}{\delmax} \right)^{\frac{1}{\zeta}} + \rR^{\frac{s}{\zeta}} \right]^{-\zeta} + 1  \right)  \equiv \rhom \left[ \delone Q(r)^{-\zeta} + 1 \right] \,,
\end{equation}
where we have defined
\begin{equation}
\label{eq:models:plmk:q}
Q(r) \equiv \left(\frac{\delone}{\delmax} \right)^{\frac{1}{\zeta}} + \left(\frac{r}{R}\right)^{\frac{s}{\zeta}} \,.
\end{equation}
The shape of this model as a function of its parameters is visualized in Fig.~\ref{fig:plmk_pars}. We note that the function's logarithmic slope,
\begin{equation}
\gamma(r) = - \left( 1 - \frac{\rhom}{\rho(r)} \right) \frac{s}{Q(r)} \rR^{\frac{s}{\zeta}}  \,,
\end{equation}
remains at a value of $-s$ only across a very narrow radial range for most parameter combinations, if at all. The full freedom of this function allows for very accurate fits, but also for a number of pathological cases where not all parameters are well constrained by the simulated profiles. We find that fixing $\zeta = 0.5$ is a compromise that works for the majority of profiles. The resulting function,
\begin{equation}
\label{eq:models:plmk2}
\rho(r) = \rhom \left( \frac{\delone}{\sqrt{(\delone / \delmax)^{2} + (r / R)^{2s}}} + 1  \right)  \,,
\end{equation}
is the infalling profile that we use throughout this paper.

\subsection{Allowed ranges of parameters}
\label{sec:models:ranges}

We impose an allowed range on each parameter to avoid extreme, unphysical values. These flat priors (in logarithmic space) are listed in Table~\ref{table:par_limits}. They are designed to be uninformative, with the exception that they enforce $\rs < \rt$. Some parameters are fixed when fitting individual halo profiles (Table~\ref{table:par_limits}). We discuss the reasons behind the chosen values in detail in Appendix~\ref{sec:app:fits:limits}. In \paperthree, we will also present the distributions of best-fit parameters for individual and stacked profiles, which could be used as a prior when fitting to observational data. The individual fits can fill out the allowed range of the parameters because not all individual profiles constrain all parameters equally well, whereas the ranges are generous for the averaged profiles.

\begin{table}
\centering
\caption{Summary of the free parameters of the fitting functions for the orbiting and infalling terms, including their allowed ranges allowed in fits. If the lower and upper bounds are the same, the parameter is held fixed. The last column lists possible modifications in fits to individual profiles. If `free', the same limits apply as for averaged profiles. See Section~\ref{sec:models:ranges} and Appendix~\ref{sec:app:fits:limits} for details.}
\label{table:par_limits}
\begin{tabular}{llccc}
\hline
Symbol & Description & Lower & Upper & Ind. \\
\hline
\multicolumn{3}{l}{\rule{-7pt}{3ex} {\bf Orbiting term}} \\
$\rhos / \rhom$ & Overdensity at scale radius & $10$ & $10^{7}$ & free \\
$\rs/\rtom$ & Scale radius & $0.01$ & $0.45$ & free \\
$\rt/\rtom$ & Truncation radius & $0.5$ & $3$ & $<10$ \\
$\alpha$ & Radial evolution of slope & $0.03$ & $0.4$ & $0.18$ \\
$\beta$ & Sharpness of truncation & $0.1$ & $10$ & $3$ \\
\hline
\multicolumn{3}{l}{\rule{-7pt}{3ex} {\bf Infalling term}} & \\
$\delone$ & Normalization at $\rtom$ & $1$ & $100$ & free \\
$s$ & Power-law slope & $0.01$ & $4$ & free \\
$\delmax$ & Central overdensity & $10$ & $2000$ & free \\
$\zeta$ & Smoothness of transition & $0.5$ & $0.5$ & $0.5$ \\
\hline
\multicolumn{3}{l}{\rule{-7pt}{3ex} {\bf Model variations}} \\
$\rho_0 / \rhom$ & Overdensity at $r = 0$ & $10$ & $10^{20}$ & --- \\
$\eta$ & Term in Model B (App.~\ref{sec:app:modelb}) & $0.1$ & $0.1$ & $0.1$ \\
\hline
\end{tabular}
\end{table}


\section{Simulations and Methods}
\label{sec:methods}

In this section, we describe our simulations and algorithms. In the interest of brevity, we do not excessively duplicate information from \paperone. We briefly summarize our simulations in Section~\ref{sec:methods:sims} and the profile data in Section~\ref{sec:methods:profiles}, largely referring the reader to \paperone. We devote more detail to our fitting procedure in Section~\ref{sec:methods:fits} and Appendix~\ref{sec:app:fits}.

\subsection{N-body Simulations}
\label{sec:methods:sims}

Our analysis is based on the \erebos suite of dissipationless $N$-body simulations \citep{diemer_14, diemer_15}, which consists of $14$ simulations of $1024^3$ dark matter particles (see Table 1 in \paperone). The suite covers different box sizes and resolutions, as well as two \LCDM and four self-similar cosmologies. The first \LCDM cosmology is that of the Bolshoi simulation \citep{klypin_11}, consistent with \wmap \citep{komatsu_11}, namely a flat \LCDM cosmology with $\Omega_{\rm m} = 0.27$, $\Omega_{\rm b} = 0.0469$, $\sigma_8 = 0.82$, and $n_{\rm s} = 0.95$. The second is a \planck-like cosmology \citep[][$\Omega_{\rm m} = 0.32$, $\Omega_{\rm b} = 0.0491$, $h = 0.67$, $\sigma_8 = 0.834$, and $n_{\rm s} = 0.9624$]{planck_14}. 

We also consider four self-similar Einstein-de Sitter universes with power-law initial spectra of slopes $n = -1$, $-1.5$, $-2$, and $-2.5$. Here, length and time are scale-free, meaning that the density profiles are independent of redshift when rescaled by a meaningful radius. The self-similar simulations highlight the impact of the initial power spectrum \citep{efstathiou_88, knollmann_08} and allow us to test a wide range of cosmologies with few simulations because \LCDM can be seen as an interpolation between different power spectrum slopes. For example, redshift trends in \LCDM profiles are mostly trends in $n$ (\paperone), meaning that a good fit of our function to profiles from self-similar cosmologies implies a wide range in redshift.

The power spectra for the \LCDM simulations were generated using \textsc{Camb} \citep{lewis_00}. They were translated into initial conditions using \textsc{2LPTic} \citep{crocce_06}. All simulations were run with \textsc{Gadget2} \citep{springel_05_gadget2}. We use the phase--space halo finder \textsc{Rockstar} \citep{behroozi_13_rockstar} to identify haloes and subhaloes. We construct merger trees by connecting halos across time using the \textsc{Consistent-Trees} code \citep{behroozi_13_trees}. The resulting halo catalogues are described in detail in \citet{diemer_20_catalogs}. 

\subsection{Dynamically split density profiles}
\label{sec:methods:profiles}

The key new ingredient on which we base our models is the splitting of dark matter particles into infalling and orbiting, the transition between which is defined to occur at a particle's first pericentre \citep[though alternative definitions exist, e.g.,][]{sugiura_20, garcia_22}. The pericentres are reliably detected by a novel algorithm that follows the orbit of each particle in each halo. This algorithm was implemented in the \sparta framework \citep{diemer_17_sparta, diemer_20_catalogs} and applied to all $14$ \erebos simulations (\paperone). Despite a few ambiguous particle orbits (e.g., due to the limited time resolution of snapshots), the resulting split profiles were shown to be robust.

We apply the same resolution cuts as in \paperone. We consider only host halos with at least $500$ particles within $\rtom$, and we cut out those parts of profiles that lie within $r_{\rm min} = {\rm max}[4\epsilon, 0.133\ \Omega_{\rm m}^{1/3} l(z)]$, where $\epsilon$ and $l(z)$ are the comoving force resolution and inter-particle spacing of the simulation. These cuts limit the effects of suppressed centripetal forces and two-body scattering \citep[see Appendix~A1 of \paperone for details;][]{power_03, ludlow_19, mansfield_21_resolution}. We also impose a limit on the fraction of a halo's mass that is unbound, $\mtomall / \mtombnd < 1.5$, which excludes halos that are being disrupted by interactions with neighbours (\paperone). However, we find that the effects of neighbours can still significantly distort the mean orbiting profiles of low-mass halos near the truncation. Thus, we exclude profile bins where $\rho < 0.1\ \rhom$ from the fits to mean orbiting profiles, although we do plot those bins in the following figures. 

We fit both individual halo profiles and averaged (mean and median) profiles. The latter are constructed by considering halo samples defined by a range of peak height (or mass), a redshift, and possibly mass accretion rate. Each sample combines halos from different \erebos simulation boxes. In the self-similar simulations, there is no physical time, so that each sample combines profiles from different redshifts (\paperone). We calculate a bootstrap uncertainty of the averaged profiles by randomly subsampling them $500$ times. 

When selecting individual halos, we again apply a lower radial limit of $r_{\rm min}$ (Section~\ref{sec:methods:profiles}) to avoid poorly resolved parts of the profiles, and we also omit bins that are expected to contain fewer than $5$ particles based on the mean profile of the given sample (\paperone). Additionally, we only fit well-resolved halos with $\ntom \geq 5000$ because their larger resolved radial range leads to less noisy distributions of the best-fit parameters. In total, we have fitted the profiles of about \num{378000} halos in the \wmap, \planck, and self-similar cosmologies. The \wmap sample shrinks with redshift from about \num{45000} at $z = 0$ to about \num{1300} at $z = 6$. Similarly, the self-similar samples shrink from about \num{120000} halos in the $n = -1$ simulation to about \num{13000} in the $n = -2.5$ box.

\subsection{Fitting procedure}
\label{sec:methods:fits}

Fitting density profiles is notoriously fickle, and the chosen routine can systematically influence the results \citep[e.g.,][]{oneil_21}. We have designed a robust routine, which we describe in detail in Appendix~\ref{sec:app:fits:proc}. We minimize the logarithmic residual between data and fitting functions using a Cauchy loss function, which severely reduces influence of outliers. For averaged profiles, the residual is compared to the bootstrap error plus a 5\% systematic uncertainty added to all bins. For individual halos, we use the sum of the Poisson error due to (potentially) low number of particles and an arbitrary 25\% systematic error. Each fit is performed multiple times from a number of initial guesses to ensure that the true minimum is found (see Appendix~\ref{sec:app:fits:proc} for details).


\section{Results}
\label{sec:results}

\begin{figure*}
\centering
\includegraphics[trim =  2mm 9mm 0mm 3mm, clip, width=\textwidth]{\figdir/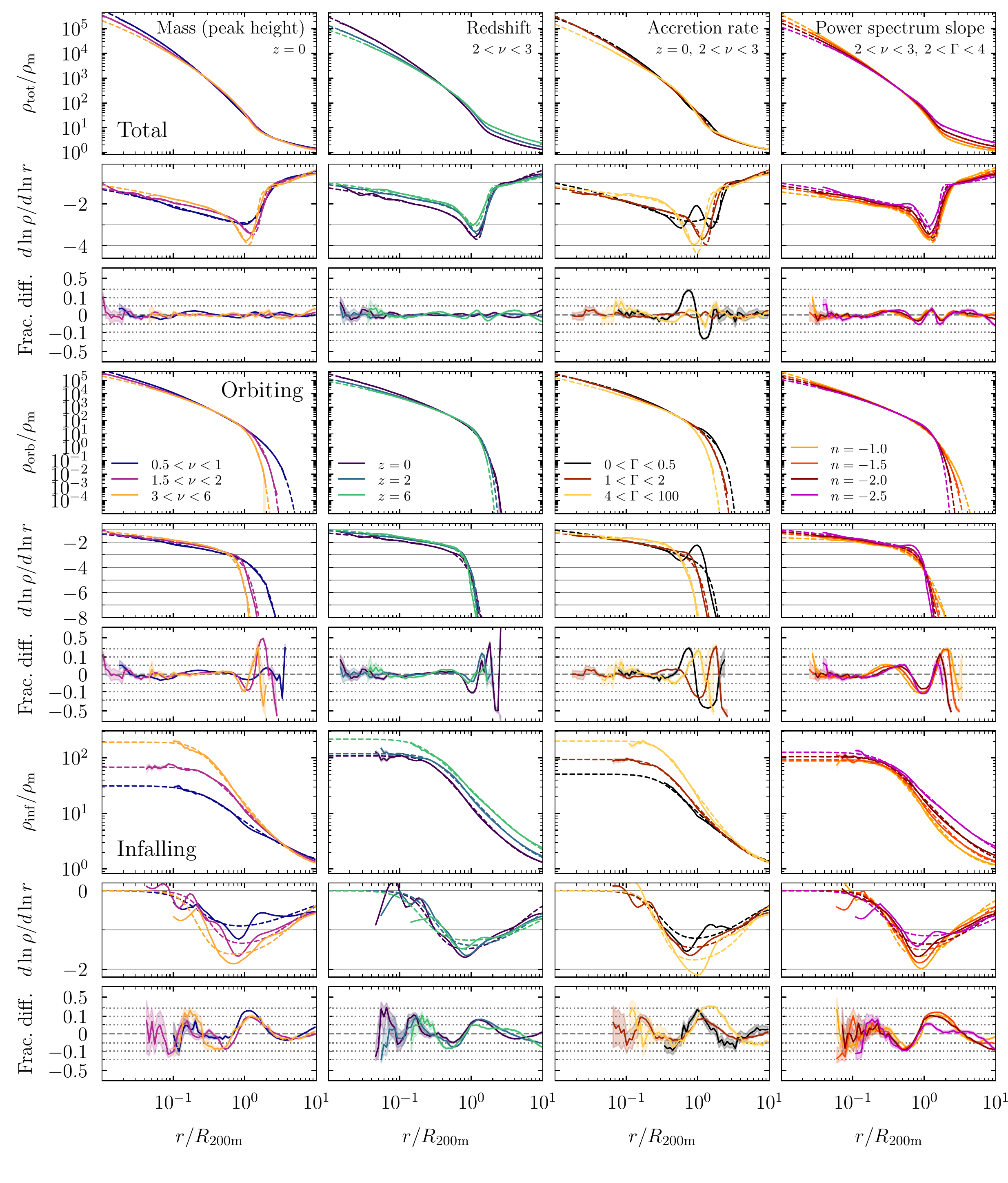}
\caption{Fits to the selected median profiles of halos binned by mass, redshift, accretion rate, and power spectrum slope (from left to right). Large panels show the total, orbiting, and infalling profiles (solid lines, from top to bottom), as well as fits (dashed lines). The smaller bottom panels show the logarithmic slope and the relative difference between simulation and fit, plotted on a symmetric log-scale where the range between $\pm5\%$ is linear; the dotted gray lines mark 5\%, 10\%, and 20\%. Shaded areas highlight the statistical uncertainty. The first column shows mass-selected samples in the \wmap cosmology. The second column shows the redshift evolution of a representative peak height bin ($2 < \nu < 3$). In the third column, this bin is additionally split by accretion rate. The fourth column compares profiles at fixed $\nu$ and $\Gamma$ in self-similar cosmologies. The total profiles are fit to 5\% accuracy, with the only exception of the wiggles in low-$\Gamma$ profiles that are caused by particles on their second orbit. The orbiting profiles are fit to $\sim 10\%$ except where they sharply drop and even tiny differences in density lead to large relative differences. The infalling profiles are similarly fit to between 10\% and 20\% accuracy, depending on the halo sample.
}
\label{fig:fits_md}
\end{figure*}

\begin{figure*}
\centering
\includegraphics[trim =  2mm 9mm 0mm 3mm, clip, width=\textwidth]{\figdir/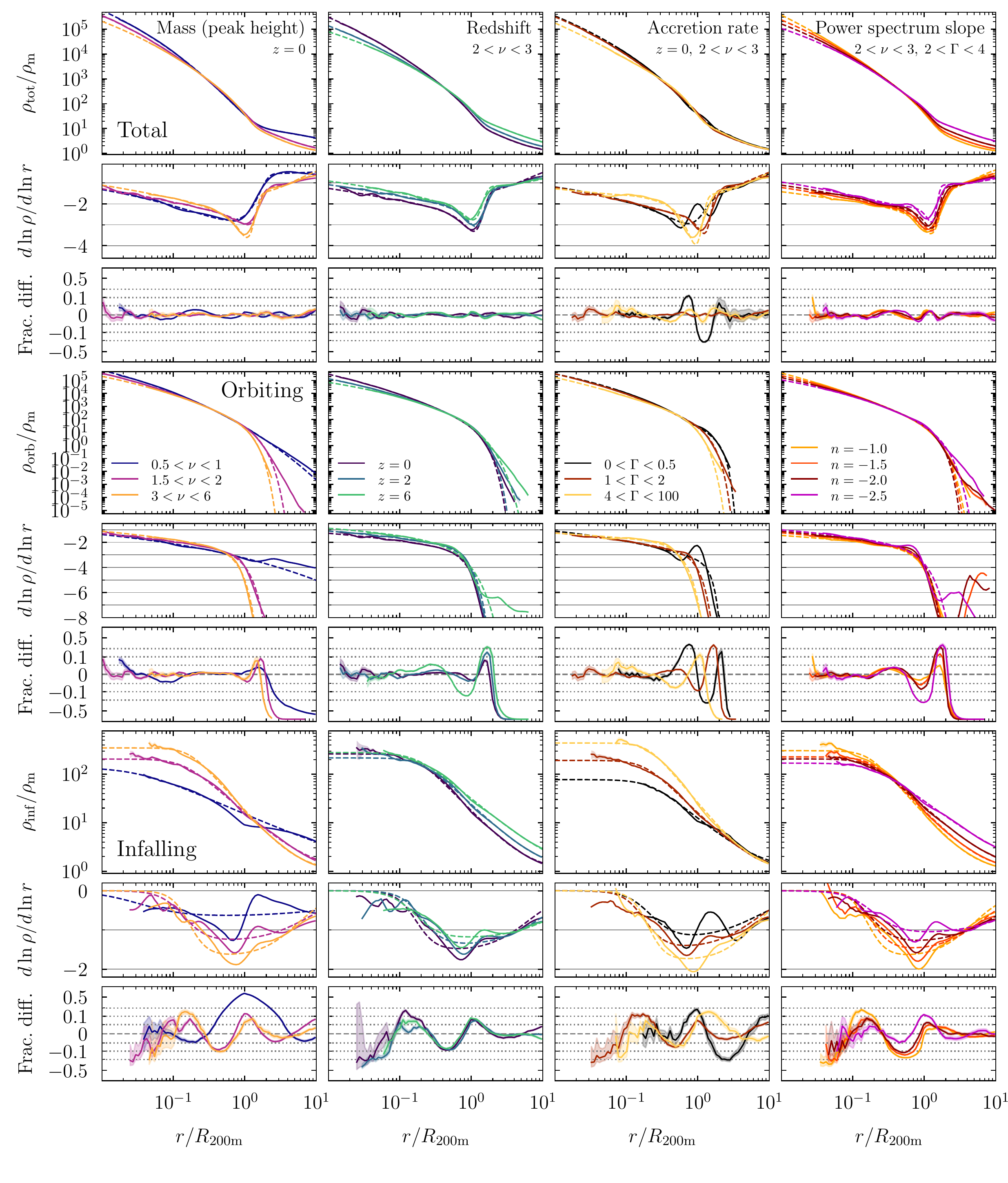}
\caption{Same as Fig.~\ref{fig:fits_md} but for mean rather than median profiles. The total mean profiles are fit as well as their median counterparts. Similarly, the mean orbiting profiles are fit to about $10\%$ accuracy except at the truncation. The fit ignores bins where the density falls below $\rho = 0.1\ \rhom$; the corresponding deviations are driven by the influence of neighbouring halos and thus not particularly meaningful. The mean infalling profiles are fit less accurately than the medians, largely because they show strong effects of the $\rtom$ boundary that is used to define subhalos. However, this issue has no impact on the total profiles.
}
\label{fig:fits_mn}
\end{figure*}

In Sections~\ref{sec:results:av} and \ref{sec:results:ind}, we fit averaged and individual profiles, respectively, and analyse the quality of the fits. Given the large parameter space of mass, redshift, cosmology, and accretion rate, we show only representative examples and refer the reader to a collection of additional online figures (Section~\ref{sec:intro}). 

\subsection{Fits to averaged profiles}
\label{sec:results:av}

Figs.~\ref{fig:fits_md} and \ref{fig:fits_mn} show median and mean profiles for a wide variety of halo samples. We omit intermediate bins to avoid crowding the figures; their profiles fall between those shown. As in \paperone, all profile plots are split into total, orbiting, and infalling profiles (from top to bottom), with smaller panels showing the logarithmic slope and the fractional deviation of the fitting function from the profiles. We show the latter on a symmetric log scale, with the region between $\pm 5\%$ being linear. The dotted lines highlight differences of $5\%$, $10\%$, and $20\%$ to guide the eye. 

In this section, we focus on fits with the full parameter freedom, that is five parameters for the orbiting term (Section~\ref{sec:models_orb:moda}) and three for the infalling term (Section~\ref{sec:models:plmk}). The orbiting and infalling profiles are fit separately with the respective models, and the results serve as initial conditions for a combined fit to the total profiles (in which we vary all parameters except for $\delmax$; Section~\ref{sec:models:ranges}). Thus, the total fits differ from the sum of the orbiting and infalling fits. In \paperthree, we will calibrate some parameters to produce a more predictive profile model. 

\subsubsection{The total profile}
\label{sec:results:av:total}

We begin by evaluating fits to the median and mean total profiles, shown in the top rows of Figs.~\ref{fig:fits_md} and \ref{fig:fits_mn}. These fits do not make use of the separate orbiting and infalling components. In the left columns, we split halo samples from the \wmap cosmology by mass, or rather by peak height, $\nu$. The corresponding mass bins range from $\mtom \approx 1.4 \times 10^{10}\ \msun$ ($\nu = 0.5$) to $\mtom > 8 \times 10^{14}\ \msun$ ($\nu > 3$). In the second column we investigate the redshift evolution of the $2 < \nu < 3$ bin, which corresponds to $1.4 \times 10^{14} < \mtom / \msun < 8 \times 10^{14}$ at $z = 0$ and to $7 \times 10^8 < \mtom / \msun < 4.2 \times 10^{10}$ at $z = 6$. The total profiles are fit excellently, to within 5\% or better. The slope panels demonstrate that the fits faithfully reproduce the complicated shape evolution of the profiles. Moreover, we do not notice any significant differences in the fit quality to the median and mean total profiles. The results are essentially the same for the \planck cosmology or for other bins in peak height.

\dkft and \paperone showed that the profile shapes depend more fundamentally on the mass accretion rate, $\Gamma$, than on mass, redshift, or cosmology. Many of the apparent trends with mass are actually trends in $\Gamma$, which are convolved with the mass-dependent distribution of accretion rates (larger halos accrete more actively when structure forms hierarchically). In the third columns of Figs.~\ref{fig:fits_md} and \ref{fig:fits_mn}, we split the $2 < \nu < 3$ bin by accretion rates; the trends are the same for other peak heights (\paperone). An accretion rate of $\Gamma \lsim 0.5$ corresponds to pure pseudo-evolution due to the evolving overdensity threshold \citep{diemer_13_pe}, meaning that the sample shown in black has undergone essentially no actual changes to the density profile over the past dynamical time. Conversely, halos with $\Gamma > 4$ (yellow) have grown rapidly. Fitting the resulting profiles is more challenging than for mass-selected samples because they exhibit a greater diversity of shapes. Nonetheless, our model fits the $\Gamma$-selected profiles almost as well as their $\nu$-selected counterparts, namely to about 5\% within the statistical uncertainties in most samples and to about 10\% for the most extreme (low or high) accretion rates. One exception are the `wiggles'  near the second caustic in low-$\Gamma$ profiles, where the fit can deviate by up to 20\%. These caustics form at the second orbit of recently accreted particles \citep{adhikari_14}, but modelling this feature in our function would introduce significant complexity for a modest return in accuracy. 

Finally, the right columns of Figs.~\ref{fig:fits_md} and \ref{fig:fits_mn} show profiles from the self-similar simulations. We have selected halos by both $\nu$ and $\Gamma$ to isolate the effect of the slope of the linear power spectrum, $n$. Even though extreme values such as $n = -1$ lead to profiles that differ significantly from \LCDM, the total profiles are fit to 5\% or better (subject to the trends with $\Gamma$ discussed above). This match gives us confidence that our model can fit profiles in a wide range of cosmologies.


\subsubsection{The orbiting term}
\label{sec:results:av:orb}

We now turn to the orbiting profiles (rows 4--6 in Figs.~\ref{fig:fits_md} and \ref{fig:fits_mn}). The exact quality of the fit is somewhat academic because the orbiting and infalling terms cannot be exactly distinguished in observations, but it was a stated goal for our model to capture the physical nature of the orbiting term. At its truncation, however, the orbiting term becomes slightly sensitive to the numerical definition of particles' pericentres (\paperone). Moreover, particles can be tidally stripped by interactions with nearby neighbours. While we try to suppress this effect by excluding halos with a large unbound component (Section~\ref{sec:methods:profiles}), a few `orbiting' particles can be dragged to arbitrarily large radii. Given that this artefact is caused by relatively few halos, the resulting differences are much more apparent in the mean than in the median profiles. These issues caution us not to over-interpret fitting errors at very low densities, which can reach $\rho = 10^{-5}\rhom$ in some mean profiles. We thus ignore densities below $\rho < 0.1\ \rhom$ in fits to the mean orbiting profiles.

The $\nu$-selected orbiting profiles are fit to 10\% accuracy out to roughly the truncation radius, where the relative differences can become arbitrarily large. As for the total profiles, we find a somewhat degraded fit quality in $\Gamma$-selected samples, chiefly because the second-caustic wiggles are more pronounced in the orbiting than in the total profiles where the infalling term partially smooths them out. Overall, the fit quality is a little better for the median profiles than for the means.


\subsubsection{The infalling term}
\label{sec:results:av:inf}

Fitting the infalling term to small radii is an entirely new challenge because the shape of this contribution at $r \lsim 0.5\ \rtom$ was hitherto concealed by the orbiting component. Key insights from \paperone were that the infalling term approaches a constant value at small radii, continuously steepens with radius towards the truncation radius, and flattens again beyond $\rt$. However, the detailed profile shapes are diverse and depend on $\nu$, $\Gamma$, and $n$, as evidenced by rows 7--9 of Figs.~\ref{fig:fits_md} and \ref{fig:fits_mn}.

Our model fits most median infalling profiles to about 10\% and some $\Gamma$-selected samples to 20\% accuracy, with no real trend with redshift, mass, or cosmology. Some of the deviations seen at small radii are not statistically significant considering the uncertainties (shaded areas). The mean infalling profiles are more difficult to fit and can deviate by more than 20\% for low-mass samples. One reason is that the mean profiles are noticeably affected by nearby neighbours, and thus by the definition of subhalos as residing within $\rtom$ (\paperone). This arbitrary boundary leads to sharp features around $\rtom$ (e.g., blue lines in the left column of Fig.~\ref{fig:fits_mn}), which cannot (and should not) be fit by our simple power-law expression.

\subsubsection{Summary}
\label{sec:results:av:summary}

We find that the new truncated-exponential model provides 5--10\% fits to the total mean and median profiles selected by mass and/or accretion rate. Any features that are systemically fit poorly are understood to be somewhat unphysical (e.g., due to neighbouring halos), too detailed for a fitting model (second caustic), or susceptible to noise (the exact shape at the orbiting truncation, where the density plummets to very low values). We note that baryonic effects enter at about the same level of accuracy, meaning that a more accurate fit would need to rely on hydrodynamical simulations. Fitting the orbiting and infalling terms separately poses a greater challenge, but the fitting function still captures their salient features and provides good accuracy over the radial ranges where the profiles are well-constrained.

We have also experimented with fitting the averaged profiles of subhaloes. Here, we select by bound (as opposed to total) mass. The \sparta algorithm keeps identifying first pericentres after a halo becomes a subhalo, but the orbiting term will increasingly include host material that the subhalo has drifted through. When binning by the resulting peak height, the truncated exponential model fits the median orbiting profiles reasonably well, as they do exhibit a clear truncation. The mean profiles, however, appear fairly irregular due to some subhaloes with strong host contributions. Similarly, the `infalling' term now contains mostly host material and follows an entirely different shape from that of host haloes. Given these difficulties, we leave an investigation of subhalo profiles to future work.

\subsection{Fits to individual halos}
\label{sec:results:ind}

\begin{figure*}
\centering
\includegraphics[trim =  4mm 23mm 3mm 5mm, clip, scale=0.62]{\figdir/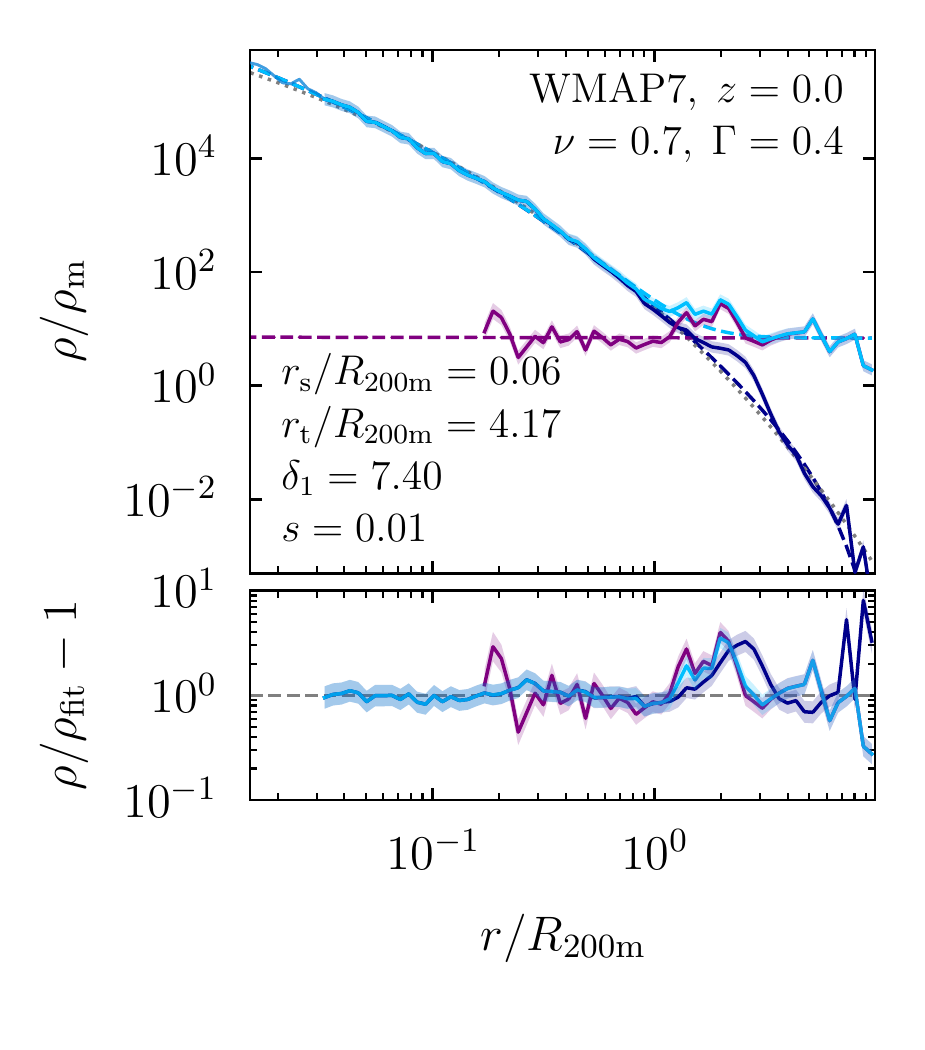}
\includegraphics[trim =  25mm 23mm 3mm 5mm, clip, scale=0.62]{\figdir/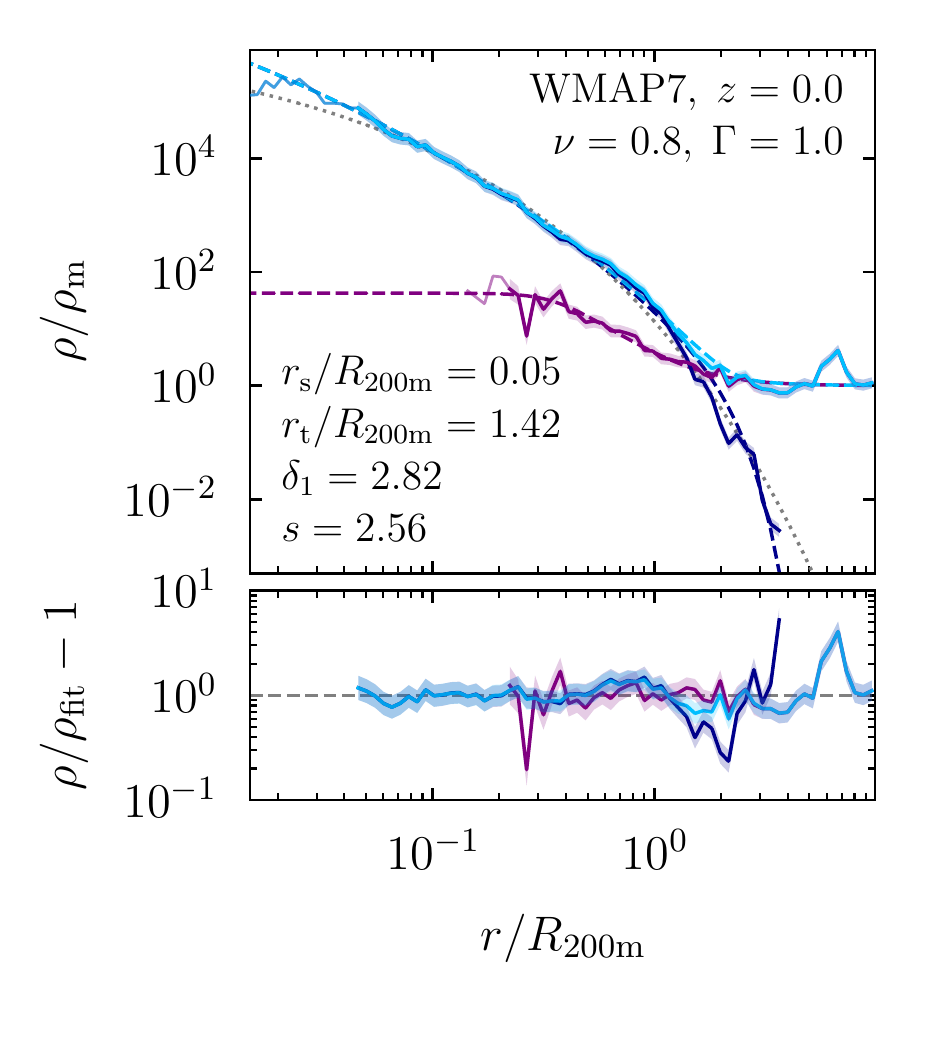}
\includegraphics[trim =  25mm 23mm 3mm 5mm, clip, scale=0.62]{\figdir/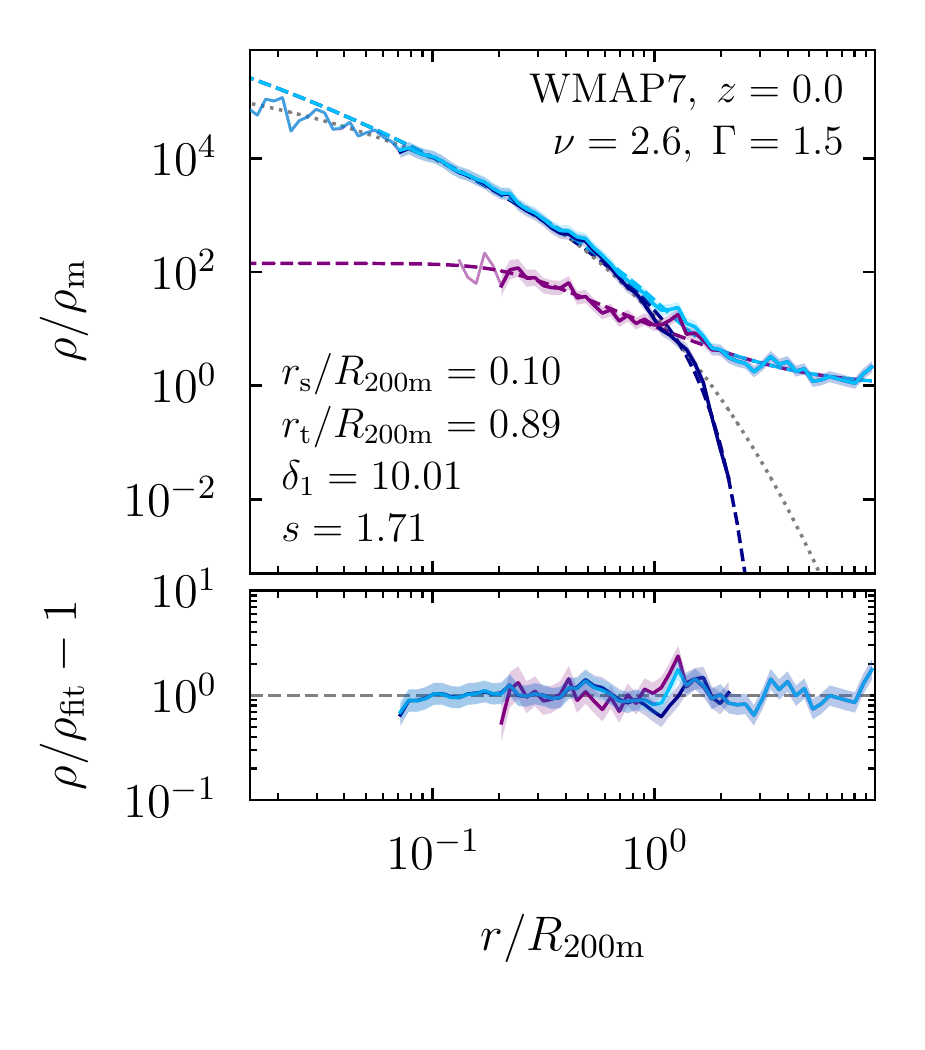}
\includegraphics[trim =  25mm 23mm 4mm 5mm, clip, scale=0.62]{\figdir/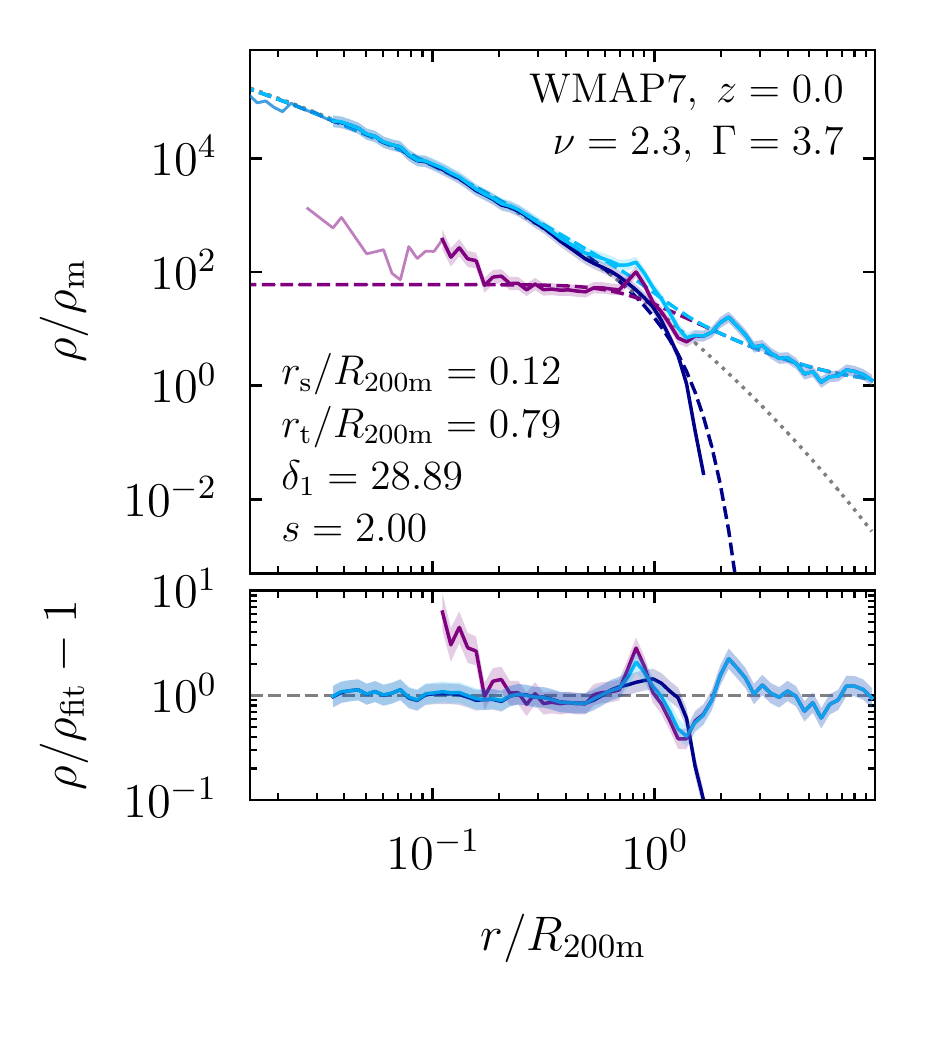}
\includegraphics[trim =  3mm 9mm 3mm 1mm, clip, scale=0.61]{\figdir/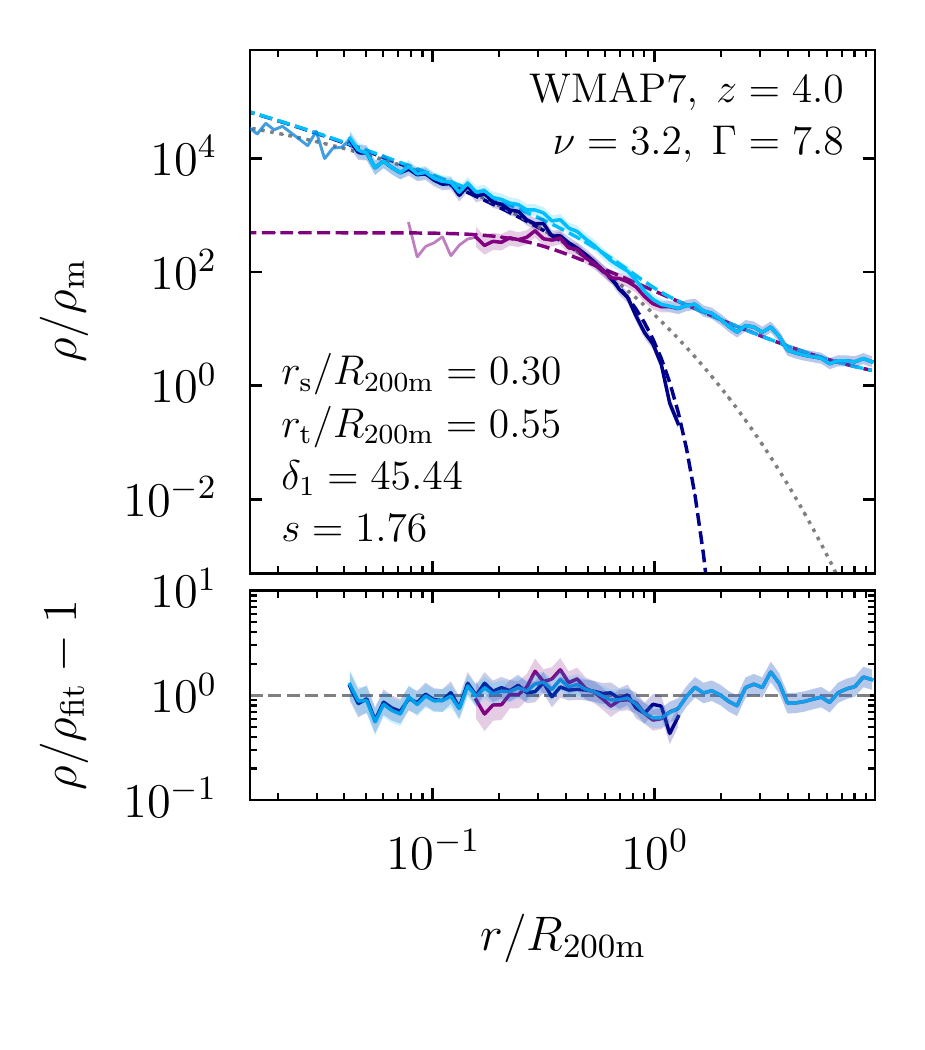}
\includegraphics[trim =  25mm 9mm 3mm 1mm, clip, scale=0.61]{\figdir/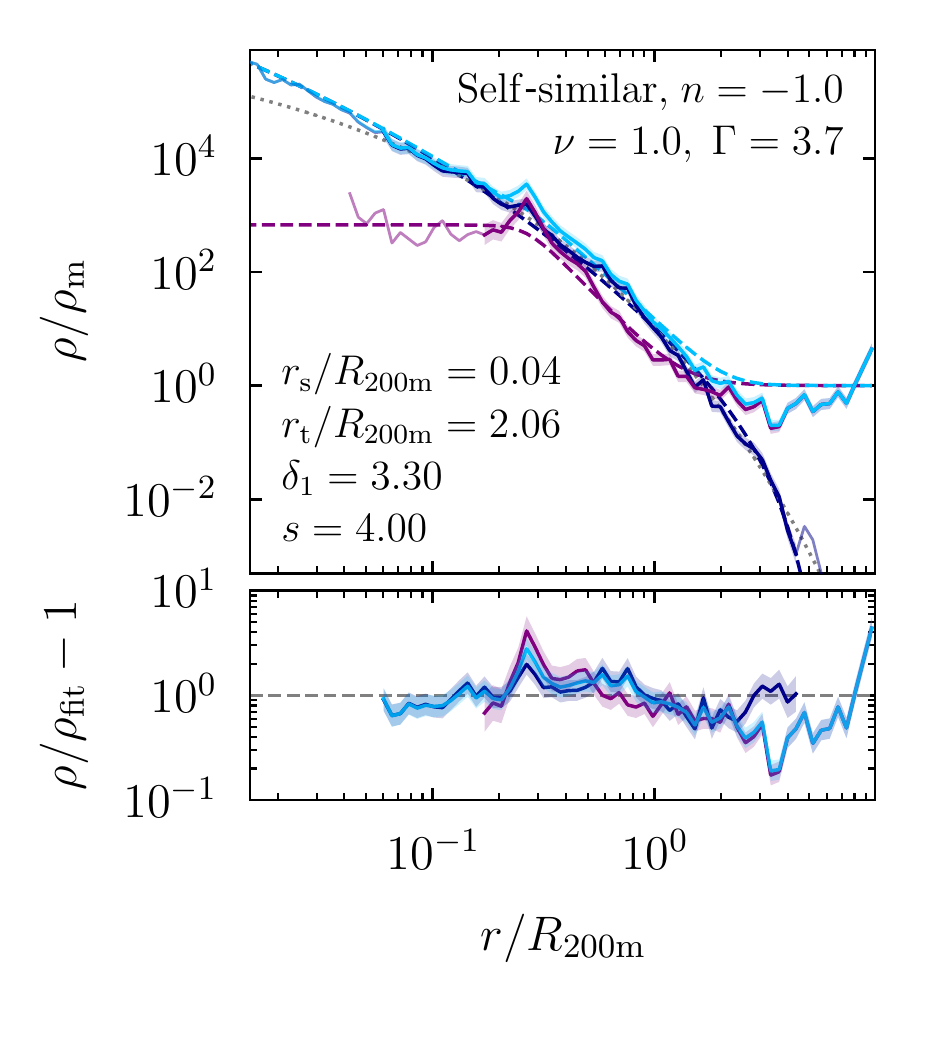}
\includegraphics[trim =  25mm 9mm 3mm 1mm, clip, scale=0.61]{\figdir/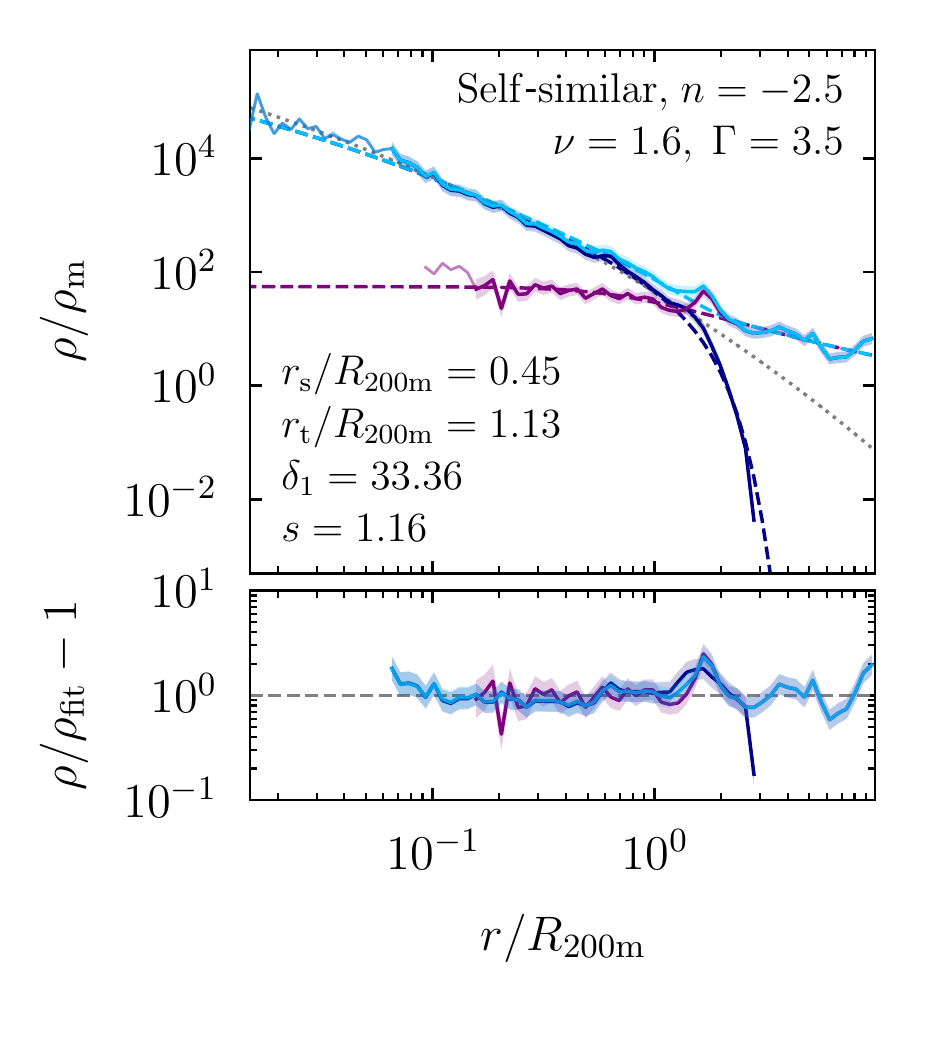}
\includegraphics[trim =  25mm 9mm 4mm 1mm, clip, scale=0.61]{\figdir/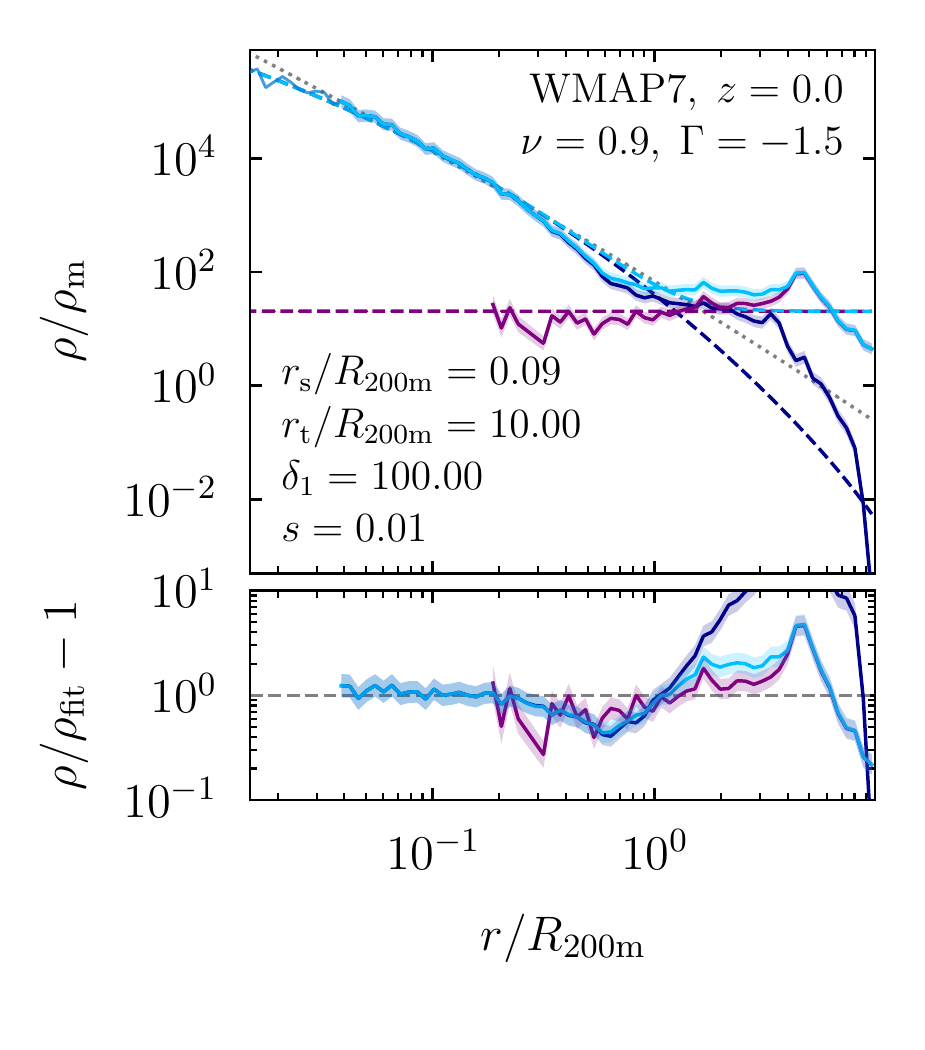}
\caption{Representative examples of fits to individual halo profiles. Solid lines show the orbiting, infalling, and total profiles (in dark-blue, purple, and light-blue, respectively). Shaded contours highlight the combination of statistical and systematic error that is used in the fits (equation~\ref{eq:syserr}). Thin lines without error contours indicate unreliable profile bins smaller than the resolution limit. The dashed dark-blue and purple lines show fits of our models for the orbiting and infalling terms; the dashed light-blue lines correspond to the sum of these fits rather than the combined fit. The dotted gray lines show a three-parameter Einasto fit to the orbiting term for comparison. The legends in each panel list properties of the halo in question and the most informative best-fit parameters. The top row shows halos from the \wmap cosmology at $z = 0$ with ascending $\Gamma$, which leads to a decreasing $\rt / \rtom$. The normalization, slope, and asymptotic values of the infalling profiles differ dramatically between individual halos. The bottom row shows a few extreme cases. First, a halo at $z = 4$ has a very high accretion rate, $\Gamma \approx 8$, which leads to a truncation radius at the lower limit of $\rt \geq 0.55\ \rtom$ as well as a high normalization of the infalling profile. The second example is a halo in the $n = -1$ self-similar cosmology, which often produces very steep infalling profiles that can reach our maximum of $s = 4$. Conversely, the $n = -2.5$ cosmology commonly produces shallow, power-law like total and infalling profiles with low concentrations (third example). The final example was deliberately picked to represent a poor fit, in this case because of a large amount of orbiting material at large radii. While an Einasto profile with the same number of free parameters ($\alpha$ instead of $\rt$) can fit low-$\Gamma$ profiles, it cannot capture the sharp truncation (dotted lines).
}
\label{fig:fits_ind}
\end{figure*}

One requirement for our fitting function was that it should fit the total profiles of individual halos, as well as their dynamically split components. For each halo in our sample of \num{378000}, we separately fit \eqmn{eq:models:moda:s2} to the orbiting term (fixing $\alpha = 0.18$ and $\beta = 3$) and \eqmn{eq:models:plmk2} to the infalling term (Section~\ref{sec:methods:fits}). We then use the best-fit parameters as the initial guess in a combined fit to the total profile. As discussed previously, individual halo profiles generally do not contain sufficient information to simultaneously constrain $\rs$ and $\alpha$, or to reliably determine $\rt$ (necessitating other ways to measure the splashback radius, such as the \sparta algorithm). Thus we experiment with different levels of parameter freedom in the combined profiles. In one case, we let all parameters except $\delmax$ float. This procedure naturally leads to the best fit in a statistical sense, but we observe numerous degeneracies and unphysical behaviour where the infalling profile adjusts to make up for a poor fit to the orbiting term. To avoid these issues, we also run fits where we vary only $\rhos$ and $\rs$ in the combined fit, meaning that $\rt$, $\delone$, $\delmax$, and $s$ are fixed to their values from the separate fits. Given the restricted freedom of the combined fits, they are generally very similar to the sum of the separate fits. The goal of the restricted fits is not to find the formally most optimal fit but to obtain meaningful best-fit parameters. We compare the fit quality of both types of fit to the NFW and Einasto models in Section~\ref{sec:comp:ind}. In \paperthree we show that the totally free fits do recover the parameters of the separate orbiting and infalling fits on average, albeit with large scatter.

Fig.~\ref{fig:fits_ind} shows a few representative example fits. Each panel shows the orbiting, infalling, and total profiles of an individual halo as solid lines; the logarithmic slopes are omitted because they are noisy for individual halos. Our model (dashed lines) fits the profiles well within statistical error and stochastic variations, except for the bottom-right panel which was selected to show a poor fit. The top row shows halos with increasing accretion rates. As discussed in \paperone, their peak height (mass) has very little influence on the profile shapes, but the increasing $\Gamma$ clearly manifests itself in decreasing values of $\rt / \rtom$ and a sharper truncation. The second row of Fig.~\ref{fig:fits_ind} shows a number of examples from other redshifts and cosmologies. First, at high redshift, the average mass accretion rate is higher, which can lead to very small truncation radii such as in the left bottom panel, where $\rt$ reaches the minimum imposed in our fits, $0.55\ \rtom$. However, we have not observed examples where the true $\rt$ is so much smaller than this limit that it would lead to a bad fit. The high accretion rate also leads to a prominent infalling profile, with $\delone \approx 45$ at $\rtom$; this example shows that our generous upper limit of $100$ is necessary. The second and third panels demonstrate that our functions for the orbiting and infalling profiles can fit extreme cases, which are often found in the self-similar simulations with very different power spectrum slopes. For example, the infalling profile in the second panel reaches the maximum slope of $s = 4$, and infalling matter contributes equally to orbiting matter all the way to $r \approx 0.2\ \rtom$. Conversely, the third panel shows a halo with flat infalling and orbiting profiles, which lead to $\rs$ reaching the maximum of $0.45\ \rtom$. Finally, the bottom right panel was selected to show a poor fit, in this case because of a major merger that deposited unusually large amount of orbiting material at large radii.

We conclude that the new truncated-exponential model captures individual profiles well, including the full range of halo properties and cosmologies investigated. In Section~\ref{sec:comp:ind}, we furthermore show that the model outperforms the Einasto function without a truncation. In Paper III, we analyse the resulting parameter distributions in detail.


\section{Comparison to other models}
\label{sec:comp}

\begin{figure}
\centering
\includegraphics[trim =  3mm 9mm 2.5mm 0mm, clip, scale=0.65]{\figdir/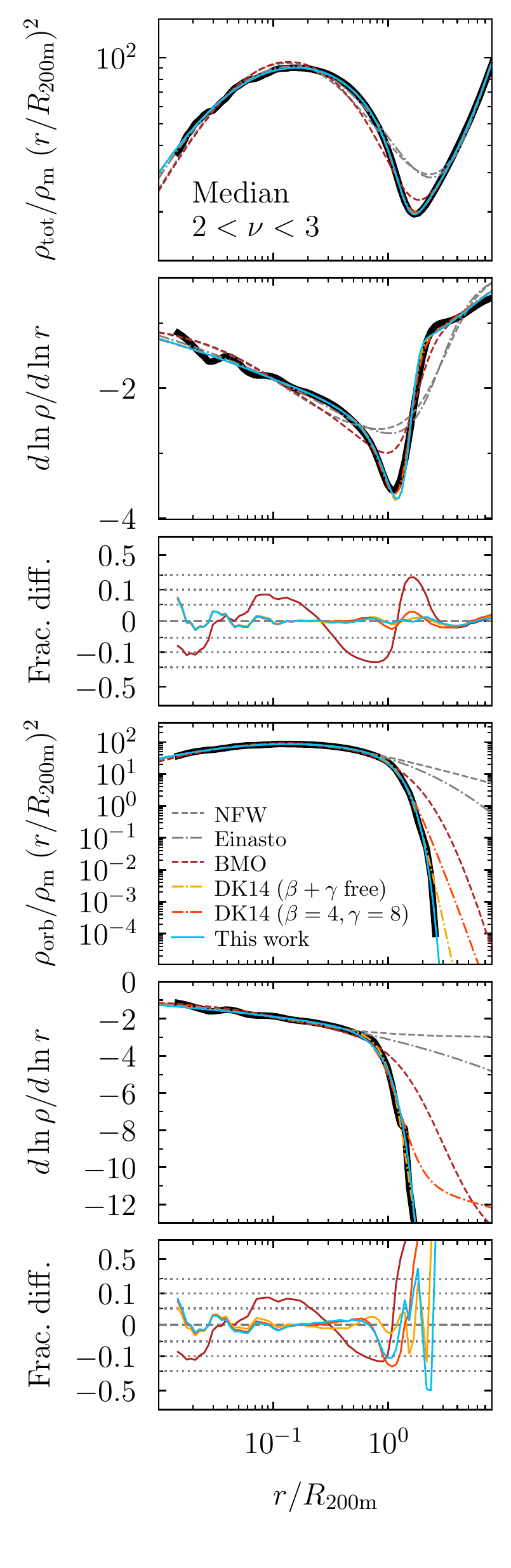}
\includegraphics[trim =  25mm 9mm 3mm 0mm, clip, scale=0.65]{\figdir/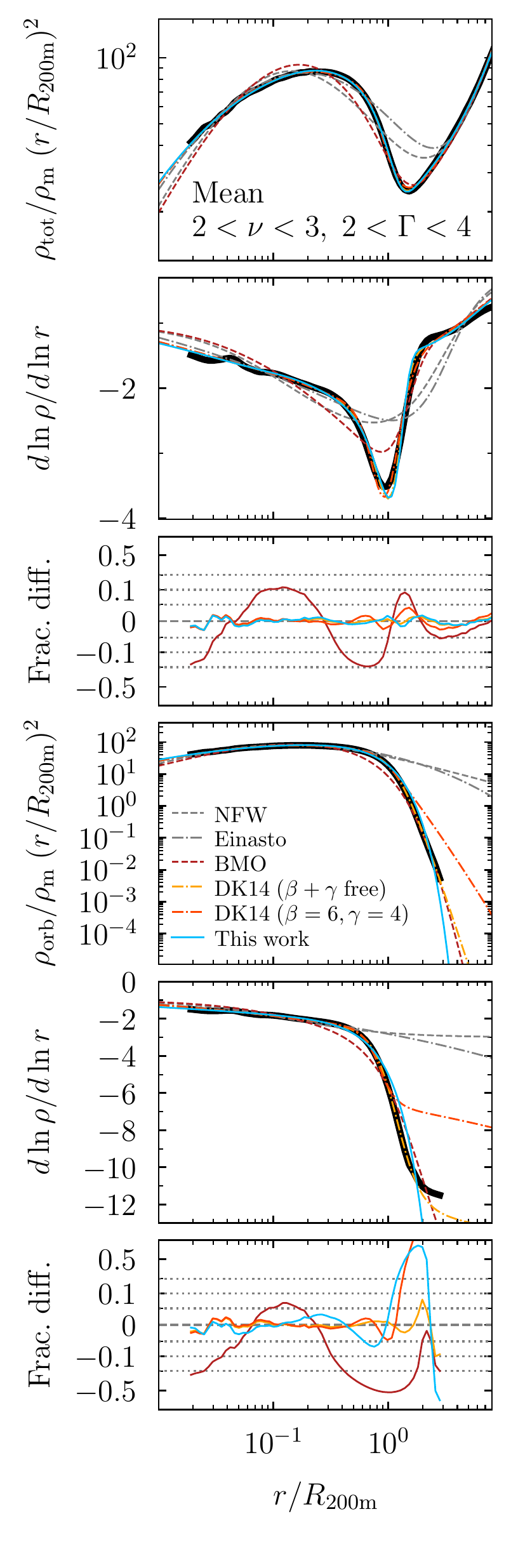}
\caption{Comparison of different fitting functions for the orbiting term. The profiles in rows 1 and 4 are rescaled by $r^2$ to compress the dynamical range. The following rows show the logarithmic slope and fractional differences, as in Figs.~\ref{fig:fits_md} and \ref{fig:fits_mn}. The columns show two representative orbiting profiles (thick black lines), namely the median profiles of halos with $2 < \nu < 3$ (left) and the mean profiles of a sub-sample additionally selected to have $2 < \Gamma < 4$ (right). The statistical uncertainty on the average profiles is smaller than the thickness of the black lines. The gray lines show NFW (dashed) and Einasto (dot-dashed) profiles, which are not designed to fit the sharp truncation of the orbiting term; those profiles are omitted from the difference panels for clarity. The coloured lines show the profiles of \citet[][BMO, dashed maroon]{baltz_09}, \dkft with the slopes $\beta$ and $\gamma$ as free parameters (yellow dot-dashed), \dkft with the slopes fixed to their suggested values for $\nu$- and $\Gamma$-selected samples (orange dot-dashed), and the new model (light blue). While the full-freedom \dkft fits well, it has an extra free parameter compared to the new function. Moreover, it is clear that its slope asymptotically approaches a certain log-linear evolution with radius due to the underlying Einasto profile. This shape fundamentally differs from the observed truncation, which is reproduced by the new model. For this figure we fit to the entire radial range of the mean profiles, including possibly unphysical features at low density that are over-fit by the free \dkft model (right column).
}
\label{fig:comp}
\end{figure}

\begin{figure*}
\centering
\includegraphics[trim =  7mm 7mm 2mm 1mm, clip, scale=0.8]{\figdir/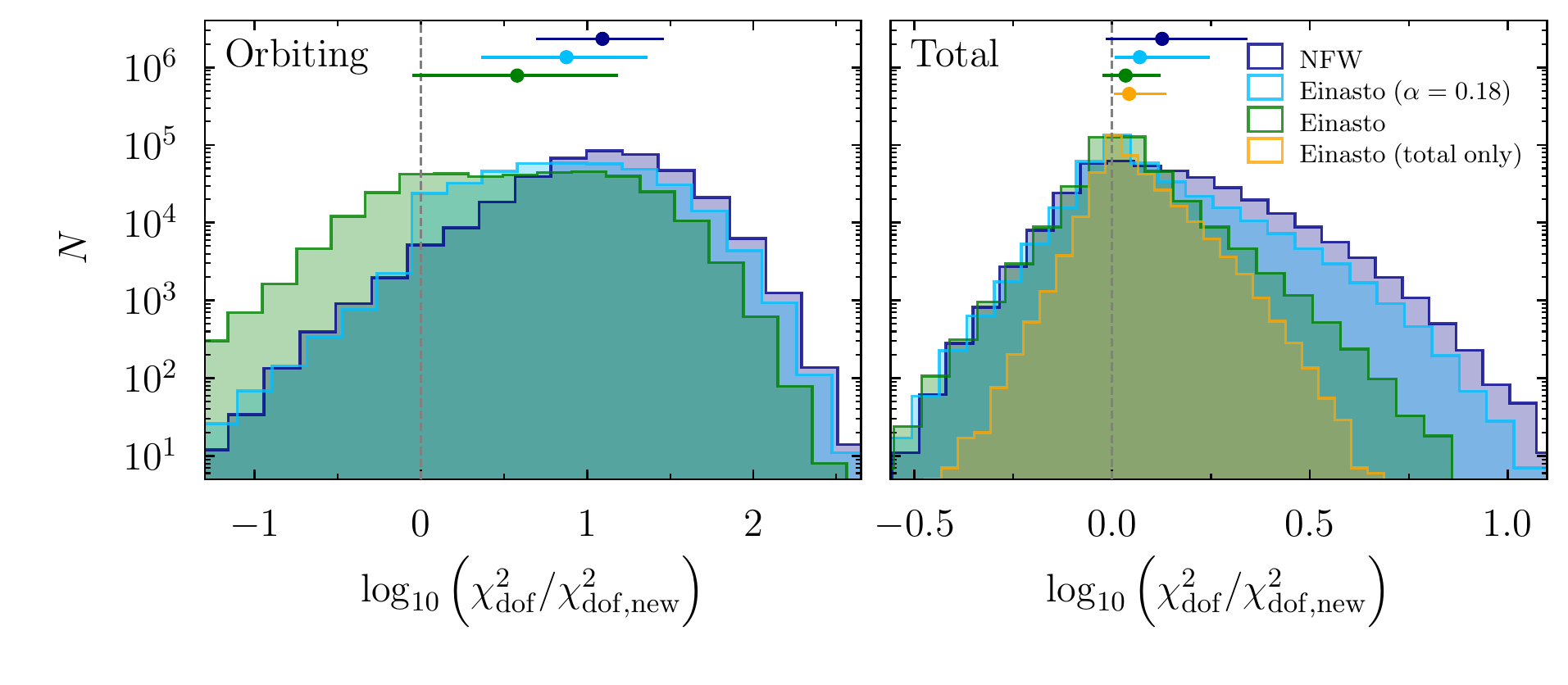}
\caption{Comparison of the relative fit quality of \eqmn{eq:models:moda:s2} and the NFW and Einasto models, for all \num{378000} halos in our individual halo sample. The functions have been applied to the orbiting term (left) and to the total profile (right). All models including the same fit to the infalling profile. The histograms show the logarithmic ratio of $\chidof$ between each model and our new model (with three free parameters, $\rhos$, $\rs$, and $\rt$). The points with error bars mark the mean ratio and 68\% interval (the median ratios are very similar). When fitting the orbiting term, the new model strongly outperforms the three-parameter Einasto profile (green), highlighting that freedom in $\alpha$ is not sufficient to fit the truncation. When fitting the total profiles but keeping $\alpha$ and $\rt$ fixed based on the separate fit (right panel), the fit quality is improved by 10\% on average (and a median of $5\%$). This result holds even when fitting only to the total term with no knowledge of the split profiles (yellow). Fixing $\alpha$ in the Einasto fits (light blue) degrades the orbiting fit but also reduces the number of degrees of freedom so that the $\chidof$ difference in the total fits remains similar. The two-parameter NFW model compares less well than the Einasto profile (dark blue).}
\label{fig:fits_ind_q}
\end{figure*}

We have convinced ourselves that the new fitting function successfully reproduces simulated profiles, but we should also ask how unique that success is and whether it could be achieved with fewer free parameters. In Section~\ref{sec:comp:av}, we compare models for the orbiting term based on averaged profiles, where we can discern the detailed shapes that lead to the success or failure of certain fitting functions. In Section~\ref{sec:comp:ind}, we evaluate the fit quality for individual halos by comparing to the NFW and Einasto forms. In Section~\ref{sec:comp:projected}, we compare the surface density profiles and lensing signal to other models and show that the features of the three-dimensional profiles do persist in projection. In Section~\ref{sec:comp:dynamical}, we consider models based on the distribution function and discuss the dynamical properties of the new model.

\subsection{Models for the orbiting term}
\label{sec:comp:av}

Fig.~\ref{fig:comp} shows a comparison of different fits to the total and orbiting profiles of two representative halo samples. The infalling part of the total profile is fit with \eqmn{eq:models:plmk2} in all cases and is thus omitted from the figure. The slope panels for the orbiting term reach down to extremely steep slopes to highlight the predictions for the asymptotic shape of this component that are made by different fitting functions (without the benefit of knowing the split profiles, of course). 

Most previously proposed fitting functions encounter at least one of two fundamental issues. First, \paperone showed that the slope of the orbiting term reaches arbitrarily steep values (as low as $-13$ in Fig.~\ref{fig:comp}). This kind of cut-off cannot be captured by power laws because they inevitably asymptote to a fixed slope at large radii. For example, the NFW profile approaches a slope of $-3$ (dashed gray lines), too shallow for the simulated profiles at $r \gsim \rt$. This issue persists for profiles with other slope transitions such as \citet{burkert_95} or superNFW \citep{lilley_18_supernfw}, models with steeper outer slopes \citep{hernquist_90}, or even more complicated combinations of power-law slopes such as generalized power-law models \citep{zhao_96, an_13, dicintio_14, dekel_17, freundlich_20}. To improve the fit of NFW profiles around the transition region, \citet[][BMO]{baltz_09} introduced a multiplicative steepening, $\rho \propto \rho_{\rm NFW} \times [\rt^2/(r^2 + \rt^2)]^n$ (maroon dashed lines in Fig.~\ref{fig:comp}). This form can fit the total profiles to about 25\% accuracy, but it fails to capture the orbiting term because it approaches a fixed slope of $-3 - 2n$. Here $n \approx 11$, a value much larger than the $n \approx 2$ usually employed to fit weak lensing signals \citep{oguri_11}. An approach similar to BMO was taken by \citet{tavio_08}; we do not include their function in Fig.~\ref{fig:comp} because it was already shown not to reach sufficiently steep slopes (fig. 15 in \dkft).

The second fundamental problem encountered by many fitting functions is that the orbiting profiles have two physical scales, which can be expressed as $\rs$ and $\rt$. These scales are set by different epochs in a halo's accretion history, namely, by the formation time, which correlates with $c$, and $\rt$, which correlates with the recent accretion history \citep{wechsler_02, tasitsiomi_04_clusterprof, zhao_09, ludlow_13, luciesmith_22_mah, shin_23}. The need for an extra variable to describe the profiles can be independently discovered using machine learning \citep{luciesmith_22_profiles}. Single-scale functions cannot address this problem and typically smooth over the sharp truncation, though they may work if the fit does not extend beyond roughly $\rvir$. The Einasto model exemplifies this behaviour: while its slope can reach arbitrarily steep values in principle, the steepening occurs at the same pace at all radii (gray dot-dashed lines in Fig.~\ref{fig:comp}). Some models have introduced a second radial scale to account for variations at small scales, e.g., in the coreNFW \citep{read_16} or coreEinasto \citep{lazar_20} models, but these modifications obviously do not change the fit at large radii. \citet{springel_99} suggested $\rho \propto \exp[-(r - \rtoc) / \rs]$ to model tidal truncation, but this term does not introduce a new scale and thus cannot capture variations in $\rt / \rs$ \citep[see also][]{fielder_20}.

To account for the second radial scale and the strong steepening, \dkft introduced a flexible truncation term (Section~\ref{sec:models:dk14}). Their results were based on the same simulations as this paper, but they did not split profiles into orbiting and infalling, meaning that their functional form (\eqmnb{eq:models:dk14}) cannot be expected to reproduce the truncation shape at low densities. Fig.~\ref{fig:comp} shows two versions of the \dkft model. In the first, the sharpness of the transition, $\beta$, and the asymptotic slope of the steepening term, $\gamma$, are free parameters (hereafter `DK14-6', yellow dot-dashed lines). In the second, they are fixed to the optimal values recommended by \dkft, namely $\beta = 4$ and $\gamma = 8$ for the $\nu$-selected sample in the left column and $\beta = 6$ and $\gamma = 4$ for $\Gamma$-selected sample on the right (hereafter `DK14-4', orange dot-dashed lines). We will keep in mind that the former version has one more free parameter than the new model, and the latter one fewer.

Fig.~\ref{fig:comp} demonstrates that the \dkft profile approaches unphysical slopes at large radii: regardless of the power-law slope $\gamma$, the exponential term from the Einasto profile eventually takes over and leads to a gradually steepening slope that does not match the simulated profiles. While this transition occurs at steep slopes (e.g., about $-10$ in the left column of Fig.~\ref{fig:comp}) and does not cause noticeable fit errors, it does lead to a conceptually wrong prediction for the shape of the orbiting term. For the purposes of this test, we have included even radii with low mean densities in the fit (unlike in our fiducial procedure, see Section~\ref{sec:methods:profiles}). 

We can get a quantitative sense of the new and \dkft fits by comparing their $\chidof$ values. The conclusions depend somewhat on whether samples are selected by $\nu$ only or also $\Gamma$, mean and median, and whether we consider the DK14-6 or DK14-4 variants. We first consider fits to only the orbiting term, which test whether the fitting function physically describe the correct underlying profile. As expected, the new form outperforms DK14-4 for virtually all mean and median samples, and often by a sizeable margin, with $\Delta_\chi \equiv \Delta \log_{10} (\chidof)$ of up to $2$. Compared to the DK14-6 form that has one more free parameter, the median $\nu$-selected profiles tend to be better fit by the new form, and the mean profiles by DK14-6. Most $\Gamma$-selected samples are better fit by DK14-6, but at the expense of  an extreme range of best-fit $\beta$ and $\gamma$. The new model better fits particularly sharp truncations in median samples with $\Gamma  \gsim 2$. All differences in fit quality are strongly reduced when fitting the total rather than the orbiting profiles, but DK14-4 still tends to perform slightly worse and DK14-6 slightly better and in most fits. Some of the latter model's success, however, arises from overfitting profiles that carry signatures of neighbouring halos, which leave a power-law tail that the \dkft fit latches onto. Physically, it is not clear that the profile model should fit such profiles.

In summary, we confirm the expected trend that models with more free parameters achieve a lower $\chidof$. However, giving the \dkft model its full freedom of $6$ parameters often leads to meaningless values of $\beta$ and $\gamma$. Our new truncated-exponential model achieves more physical asymptotic profiles and a comparable fit quality with fewer, well-defined parameters. 

\subsection{Individual profiles}
\label{sec:comp:ind}

While the examples of individual halos in Fig.~\ref{fig:fits_ind} are encouraging, we need a quantitative metric to decide whether the additional complexity of the new model is warranted compared to simpler functions. Without an absolute measure of fit quality in the absence of a well-motivated uncertainty on the profiles (Appendix~\ref{sec:app:fits:proc}), our best option is to compare the relative $\chidof$ of different fits. To this end, we repeat the procedure described in Section~\ref{sec:results:ind} with three models: NFW, Einasto with free $\alpha$, and Einasto with $\alpha = 0.18$. In Fig.~\ref{fig:fits_ind_q} we compare these fits for all \num{378000} halos in our individual sample. While there are mild trends when we break up the sample by cosmology and redshift, the overall picture remains unchanged. 

The most `fair' comparison is between the new model and an Einasto profile with free $\alpha$ because both have three parameters. When fitting only the orbiting profile, the new model has a lower $\chidof$ by a median factor of four (green histogram in the left panel of Fig.~\ref{fig:fits_ind_q}). This dramatic improvement highlights that the truncation cannot generally be fit with Einasto profiles. This observation is also apparent in Fig.~\ref{fig:fits_ind}, where low-$\Gamma$ halos are well described by the Einasto form (gray dotted lines) but where the increasingly sharp truncation leads to poor fits at large radii. 

In realistic applications, however, we fit to total profiles. This case is shown in the right panel of Fig.~\ref{fig:fits_ind_q}. We test two variations of the fits. First we vary only $\rhos$ and $\rs$ in the combined fit, keeping $\alpha$, $\rt$, and the parameters of the infalling profile to their values from the separate fits (green). The new model now improves $\chidof$ by about 10\% on average because exact shape of the truncation is less important in total profiles. Moreover, in most real-world applications, we do not have access to the separate orbiting profile. This result holds when we ignore the results of the separate orbiting fit and let all free parameters of the orbiting term vary (yellow). This relatively modest difference reminds us that it is difficult to extract the truncation radius from total individual profiles, but it still represents a measurable improvement over the Einasto model. We also compare the new model to the two-parameter Einasto fit with $\alpha = 0.18$ and to the NFW profile (light and dark blue histograms in Fig.~\ref{fig:fits_ind_q}). The orbiting profile is described much less accurately by these forms, with mean $\chidof$ increases of about $7$ and $12$, respectively. Once again, the difference is much smaller when fitting the entire profile, with mean increases of 17\% and 34\%. We have confirmed that these results remain unchanged when directly optimizing $\chi^2$ instead of the Cauchy loss function $\ln(1 + \chi^2)$ (Appendix~\ref{sec:app:fits:proc}).

Overall, Fig.~\ref{fig:fits_ind_q} provides strong justification for our new fitting function because it improves the fit quality at a fixed number of free parameters. While this improvement is most noticeable when fitting the orbiting term separately, we also record a statistically significant improvement when fitting the total profiles without any knowledge of the orbiting term. We conclude that the new function captures both averaged and individual profiles well. In \paperthree, we analyse the connection between the properties of individual halos and their best-fit parameters.

\subsection{Projected profiles and lensing signal}
\label{sec:comp:projected}

\begin{figure}
\centering
\includegraphics[trim =  3mm 9mm 10mm 7mm, clip, scale=0.79]{\figdir/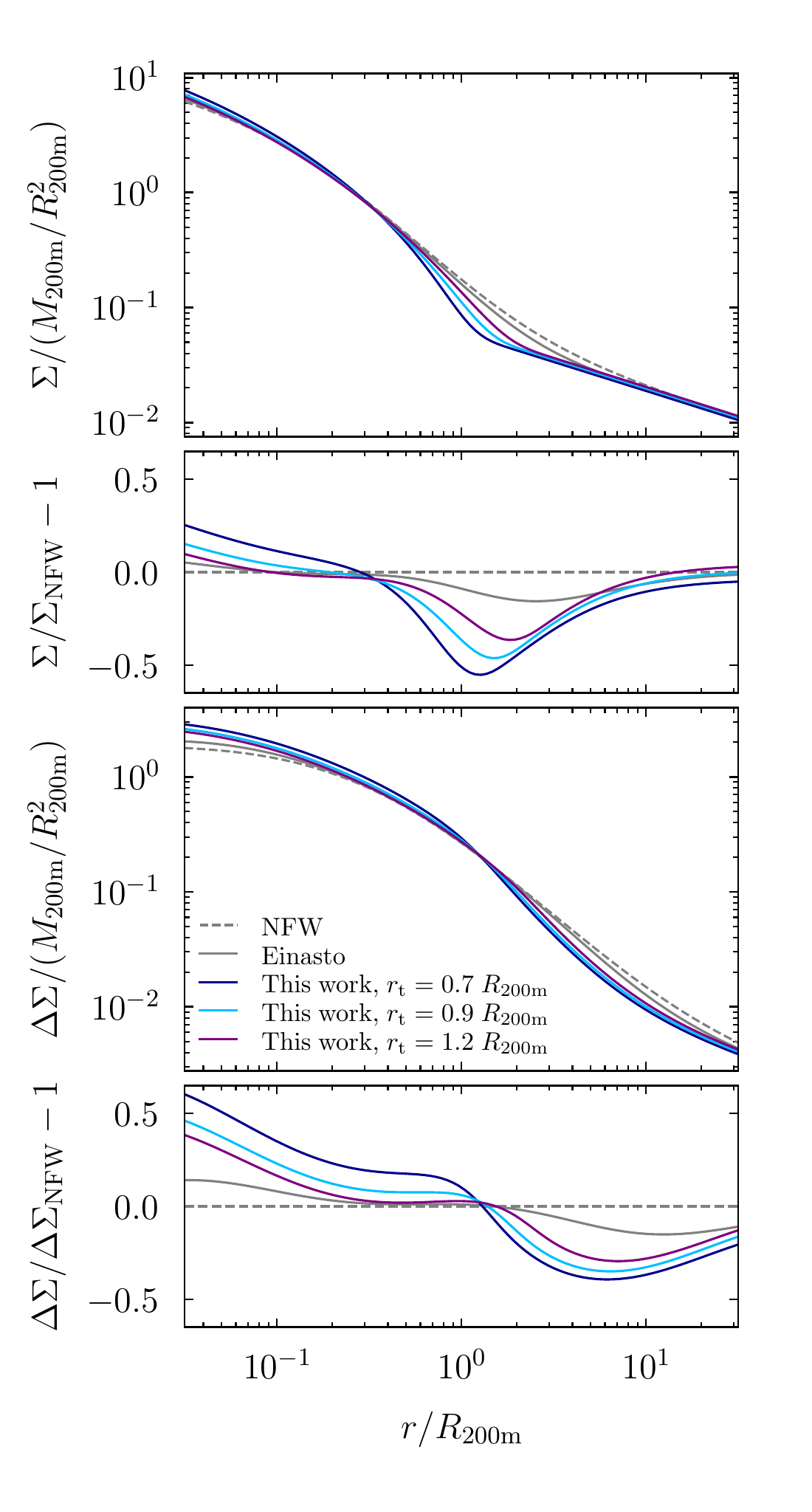}
\caption{The projected (surface) density profile (top) and lensing signal (excess surface mass density, bottom) for characteristic profiles of a cluster halo with $\mtom = 5 \times 10^{14} \msunh$ and $\ctom = 5$. We compare the NFW, Einasto, and truncated exponential profiles, where we force the truncation radius of the latter to the values listed in the legend and renormalize the profile to the same total mass. The differences in the transition region clearly propagate to the surface density profiles, where the different values of $\rt$ are easily distinguishable by eye. The differences are smaller but propagate to a much larger radial range in $\Delta \Sigma$ because $\Sigma(r)$ is subtracted from the averaged surface density.}
\label{fig:projected}
\end{figure}

While we have demonstrated that the new profile model is preferred by simulated halo profiles, the three-dimensional density profile is generally not accessible in observations. In general, we measure either the surface density or the lensing signal. The surface (or projected) density, $\Sigma(r)$, can be measured via the satellite distribution, for example \citep{more_16}. The lensing signal corresponds to the excess surface mass density, $\Delta \Sigma \equiv \overline{\Sigma}(<r) - \Sigma(r)$, the difference between the averaged surface density within a given radius and the surface density at that radius. In this section, we investigate how different the new profile is from other models in projection, and whether variations in the truncation radius can, in principle, be extracted from observational data.

In Fig.~\ref{fig:projected}, we show projected profile quantities for a typical cluster halo with $\mtom = 5 \times 10^{14}\ \msunh$ and $\ctom = 5$ at $z = 0$. We set $\alpha = 0.18$ and $\beta = 0.3$, but we vary $\rt$ between values corresponding to low and high accretion rates (about $0.7 < \rt/\rtom < 1.2$, \paperthree). The surface density profiles of these models can easily be distinguished by eye, and all truncated exponential profiles visibly differ from the NFW and Einasto forms. The differences are smaller in the lensing signal, but they persist across a wide radial range. This finding highlights that the entire profile must be fitted in order to extract the full information content \citep[e.g.,][]{xhakaj_20, xhakaj_22}.

\subsection{Dynamical models}
\label{sec:comp:dynamical}

\begin{figure}
\centering
\includegraphics[trim =  3mm 6mm 1mm 2mm, clip, scale=0.78]{\figdir/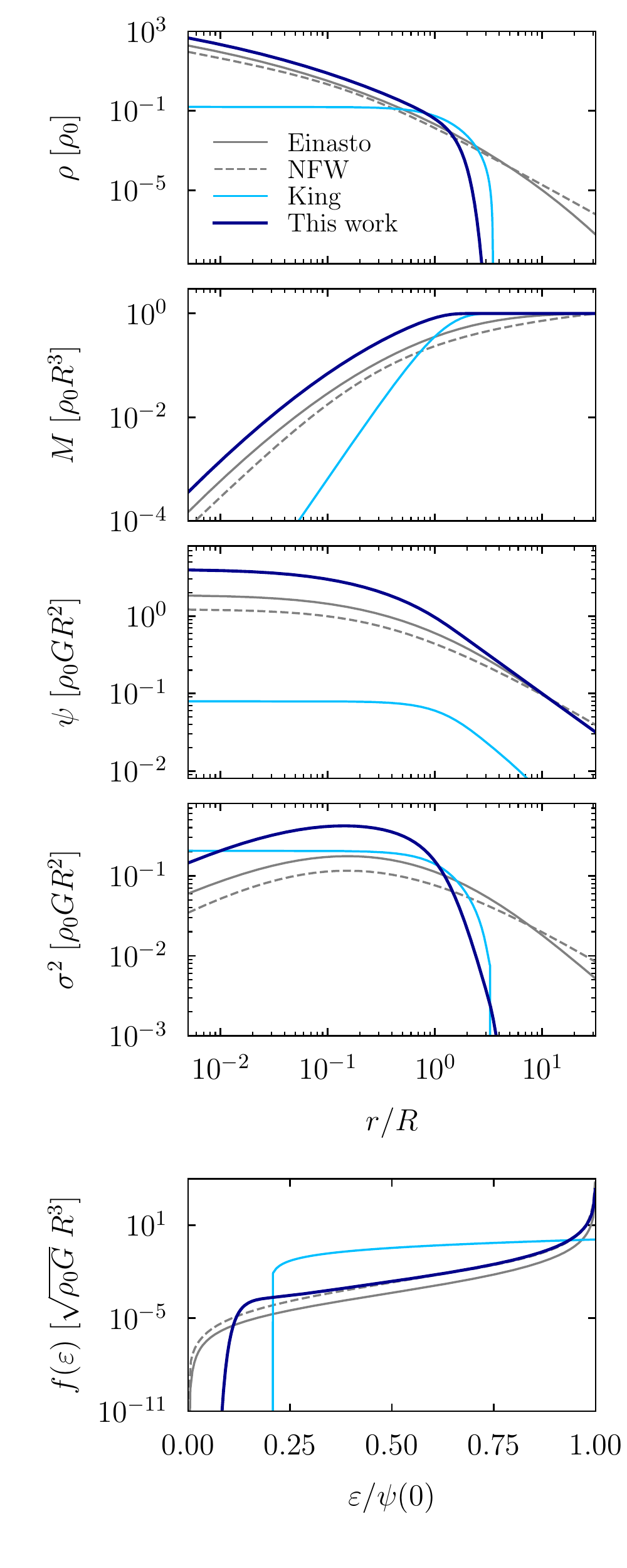}
\caption{Dynamical properties of our new model (dark blue), a \citet{king_66} profile (light blue, with a comparable cut-off radius), as well as the NFW and Einasto forms (gray). The panels show the density $\rho$, enclosed mass $M$, potential $\psi$, isotropic velocity dispersion $\sigma^2$, and distribution function $f(\varepsilon)$ (from top to bottom). The latter falls to zero for unbound particles with $\varepsilon < 0$; conversely, $\varepsilon = \psi(0)$ corresponds to maximally bound particles at the centre of the halo. The profiles are expressed in dimensionless units and normalized to reach the same total mass $M = 1$ at the right edge of the plot. Compared to the conventional models, the sharp density cut-off in the new model leads to a mass distribution that suddenly flattens, a rapidly dropping velocity dispersion, and a sharp drop in the distribution function at the energy corresponding to orbits that reach their apocentre near the truncation radius.}
\label{fig:dynamical}
\end{figure}

Assuming that halos are in dynamical equilibrium and that the particle velocities are isotropic \citep[which is not exactly realistic, e.g.,][]{hansen_06}, a model for the density profile implies a particular form of the velocity dispersion, $\sigma^2$, and the phase-space distribution function, $f(\varepsilon)$. The mathematical expressions for these quantities are given in Appendix~\ref{sec:app:comp_df}. The velocity dispersion can be measured in principle, e.g., from the motions of satellite galaxies or stars \citep{mamon_10, okumura_18, adhikari_19, hamabata_19, tomooka_20, bose_21, aung_21_phasespace, aung_22}. We do not expect the dynamics to be fundamentally altered by the presence of baryons \citep{callingham_20}.

In this section, we check whether the new model makes reasonable predictions for the dynamical quantities and we question whether other, energy-based models could make similar predictions. The former is not automatically guaranteed since not all profiles have positive, continuous distribution functions \citep{baes_21}. We define $\psi(r)$ to be the (positive) gravitational potential of the halo and $\varepsilon \equiv \psi(r) - v^2 / 2$ to be the relative binding energy of particles. Particles with $\varepsilon < 0$ would be unbound and thus lead to $f(\varepsilon < 0) = 0$. Conversely, the maximum binding energy would correspond to particles that reside at the centre of the halo with velocity $v = 0$, meaning that $\varepsilon = \psi(0)$. 

Fig.~\ref{fig:dynamical} shows radial profiles of density, enclosed mass, potential, and velocity dispersion, as well as the distribution function. We compare our new model (dark blue) to the NFW and Einasto forms (gray). The profiles are normalized to have the same total mass at the largest radius shown, and they are expressed in dimensionless units rescaled by $R$, some scale density $\rho_0$, and the gravitational constant $G$. The results thus depend only on the relative scales $\rs / R$ and $\rt / R$, as well as the steepening parameters $\alpha$ and $\beta$. We choose a representative set of parameters for all models, namely $\rs / R = 0.2$, $\alpha = 0.18$, $\rt / R = 1$, and $\beta = 4$. We observe that $M(r)$ flattens as a consequence of the truncation in $\rho$, and that the potential approaches $\psi \propto 1/r$ at a smaller radius than for the Einasto profile. 

The density cut-off is also mirrored in energy space, where the velocity dispersion falls sharply. The corresponding distribution function is similar to the NFW and Einasto profile at high binding energies, but it drops steeply at the binding energy corresponding to particles at the truncation radius. This is the desired behaviour for a profile that attempts to model the edge of the orbital distribution \citep[e.g.,][]{king_66, drakos_17, amorisco_22}. In other words, the splashback radius (the apocentre of the most recently accreted particles) is governed by a soft limit in the distribution of particle kinetic energies. Conversely, the NFW and Einasto profile have small, but finite, support out to the smallest binding energies \citep[see also, e.g.,][]{cardone_05, mamon_05_a, beraldoesilve_14, baes_19}. For large values of $\beta$ such as $10$ (i.e., very sharp truncation in density space), the distribution function can become non-monotonic near its cutoff.

These observations raise the questions of whether the truncation of the orbiting term could be described by a truncation of $f(\varepsilon)$ at some maximum energy. One well-known example of such models is the \citet{king_66} family of profiles, which corresponds to an isothermal sphere that is sharply cut off at a certain binding energy \citep[see also][]{michie_63}. This model is defined by the distribution function
\begin{equation}
\label{eq:king}
f_{\rm king}(\varepsilon) = f_0 \left( e^{(\varepsilon - \varepsilon_{\rm t}) / \sigma^2} - 1 \right) \,,
\end{equation}
where $f_0$ is a normalization, $\sigma$ the velocity dispersion of the isothermal sphere, and $\varepsilon_{\rm t}$ the energy where the object is truncated ($f \rightarrow 0$). We use the dimensionless variables of \citet{drakos_17} to numerically integrate \eqmn{eq:king} to find the potential, density, and velocity dispersion profiles. Given our fundamental scale $R$ and the mass normalization, the profile has only one free parameter $Z_{\rm t} \equiv \varepsilon_{\rm t} / [\sigma^2 \psi(0)]$. The light-blue lines in Fig.~\ref{fig:dynamical} show a King profile with $Z_{\rm t} = 0.2$, a cut-off energy chosen to roughly match the real-space truncation radius of $\rt = 0.2\ R$. The position and sharpness of the truncation can be adjusted using $Z_{\rm t}$ \citep[e.g., fig. 4.8 in][]{binney_08}, but Fig.~\ref{fig:dynamical} immediately explains why the King model cannot describe realistic halo profiles: the density approaches a large, fixed-density core at $r \ll R$. Moreover, the truncation in binding energy leads to a sharper real-space truncation than we observe in simulations. Our new model allows for lower binding energies (bottom panel) and thus for a smoother truncation.
 
Of course, the King model is only one example of dynamics-based profiles. Another recently suggested model is DARKexp \citep{hjorth_10_darkexp1}, but this model cannot fit the truncation because its slope approaches $\gamma = -4$ at large radii, regardless of the $\phi_0$ parameter that controls the distribution of particle energies \citep{williams_10_darkexp2, williams_10_darkexp3}. Similarly, \citet{pontzen_13} derive a distribution function by maximising the entropy, but it is not clear whether this model produces a truncation at the right radius (their fig.~7; see also \citealt{wagner_20}).


\section{Conclusions}
\label{sec:conclusion}

We have presented a new fitting function for halo density profiles, which is composed of models for the orbiting (one-halo) and infalling terms. The orbiting term is modelled as an extension of the Einasto profile, $\rhoorb \propto \exp \left[-2/\alpha\ (r / \rs)^\alpha - 1/\beta\ (r / \rt)^\beta \right]$. The infalling term is a power law in density that smoothly approaches a maximum value at small radii. These functions are implemented in the publicly available \colossus code, including numerical routines for their mass, surface density, and $\Delta \Sigma$ integrals. Our main conclusions are as follows. 

\begin{enumerate}

\item The new model fits mean and median total profiles to roughly $\pm 5\%$ across a vast range of radius, halo mass, redshift, and cosmology. When halos are also selected by accretion rate, the fit quality can degrade to $\pm 20\%$ near the transition between the orbiting and infalling profiles.

\item The new model accurately fits the orbiting and infalling terms separately, even down to densities well below the cosmic mean. The exact fit quality depends on the chosen mass, accretion rate, and whether mean or median profiles are considered. 

\item By fixing $\alpha = 0.18$ and $\beta = 3$, the orbiting model becomes a 3-parameter fit that captures the profiles of individual halos more accurately than Einasto profiles (on average). 

\item The sharp truncation of the orbiting term corresponds to a truncation in the binding energy of particles, which is not replicated in more extended profile models.

\item Different truncation radii lead to clear differences in the projected surface density profiles and in the lensing signal $\Delta \Sigma$.

\item We introduce an augmented `Model B' that fixes the logarithmic slope at the scale radius to be $-2$. This formulation is almost equivalent to the fiducial model but alleviates some rare parameter degeneracies at the cost of slightly increased complexity.

\end{enumerate}

In \paperthree, we will analyse the best-fit parameters of averaged and individual halo profiles. We have left a number of theoretical and practical questions unexplored. For example, we intend to provide analytical approximations for the mass and projected density of the new fitting function. Another urgent question is the relationship between the profile parameters and definitions of the halo boundary based on the orbiting population of subhalos \citep{aung_21_phasespace, garcia_21}, or other definitions based on the interplay between the infalling and orbiting components \citep{fong_21}. Finally, we intend to apply the new model to observational data.


\section*{Acknowledgements}

I am grateful to Nicole Drakos for help with computing King profiles and to Maarten Baes for verifying the dynamical properties of the new model. I thank Andrew Hearin and Keiichi Umetsu for comments on a draft, and Susmita Adhikari, Han Aung, Barun Dhar, Rafael Garcia, Daisuke Nagai, Eduardo Rozo, and Angus Wright for productive conversations. This work was partially completed during the Coronavirus lockdown and would not have been possible without the essential workers who did not enjoy the privilege of working from the safety of their homes. The computations were run on the \textsc{Midway} computing cluster provided by the University of Chicago Research Computing Center and on the DeepThought2 cluster at the University of Maryland. This research extensively used the python packages \textsc{Numpy} \citep{code_numpy2}, \textsc{Scipy} \citep{code_scipy}, \textsc{Matplotlib} \citep{code_matplotlib}, and \colossus \citep{diemer_18_colossus}.


\section*{Data Availability}

The \sparta code that was used to extract the dynamically split density profiles from our simulations is publicly available in a BitBucket repository, \href{https://bitbucket.org/bdiemer/sparta}{bitbucket.org/bdiemer/sparta}. An extensive online documentation can be found at \href{https://bdiemer.bitbucket.io/sparta/}{bdiemer.bitbucket.io/sparta}. The \sparta output files (one file per simulation) are available in an hdf5 format at \href{http://erebos.astro.umd.edu/erebos/sparta}{erebos.astro.umd.edu/erebos/sparta}. A Python module to read these files is included in the \sparta code. Additional figures are provided online on the author's website at \href{http://www.benediktdiemer.com/data/}{benediktdiemer.com/data}. The full particle data for the \erebos $N$-body simulations are too large to be permanently hosted online, but they are available upon request. 


\bibliographystyle{\includedir/citestyle_mnras}
\bibliography{\includedir/bib_mine.bib,\includedir/bib_general.bib,\includedir/bib_structure.bib,\includedir/bib_galaxies.bib,\includedir/bib_clusters.bib}


\appendix

\section{Details on fitting procedure}
\label{sec:app:fits}

In this Appendix, we give additional detail on the technical aspects of our profile fits. In Section~\ref{sec:app:fits:proc} we discuss our fitting procedure and in Section~\ref{sec:app:fits:limits} we describe the reasoning behind and meaning of the parameter limits given in Section~\ref{sec:models:ranges} and Table~\ref{table:par_limits}.

\subsection{Fitting procedure}
\label{sec:app:fits:proc}

A first important choice is the definition of the residual. The conventional approach of minimizing the square of $(\rho_{\rm fit} - \rho_{\rm data}) / \sigma$ works poorly in the outer profiles because the summed $\chi^2$ is dominated by the much higher densities at small radii. Some authors have chosen to minimize $r^2 (\rho_{\rm fit} - \rho_{\rm data}) / \sigma$ instead, but this function still represents an arbitrary weighting with radius. One can even fit the slope profile, but the results depend on how the slope is estimated from noisy data \citep{oneil_21}. To obtain a fit of equal quality at all radii, we minimize the square logarithmic differences between simulated and fitted profiles. We find that we obtain the most physically sensible fits if we aggressively reduce the effect of outliers by using a Cauchy loss function, so that the minimized quantity is 
\begin{equation}
\label{eq:chi2}
\chi_{\rm cauchy}^2 \equiv \sum_i \ln \left(1 + \chi_{\rmi}^2 \right) \quad \mathrm{with} \quad \chi_{\rmi} \equiv \frac{\ln \rho_{\rm i,fit} - \ln \rho_{\rm i,data}}{\ln(1 + \sigma_\rmi / \rho_{\rm i,data})} \,,
\end{equation}
where $\rho_{\rm i,fit}$ is the fit density in bin $i$, $\rho_{\rm i,data}$ is the density of the simulated profile (individual or averaged), and $\sigma_\rmi$ is the uncertainty in $\rho_{\rm i,data}$. Throughout the paper, we quote the usual values of $\chi^2 = \sum \chi_\rmi^2$, i.e., the logarithmic differences without taking the loss function into account. The additional logarithm in the Cauchy loss function allows the fitter to ignore features such as wiggles in favour of a better fit to the well-defined inner profile.

The next decision is how to define the uncertainty $\sigma_\rmi$ of each profile bin. The bootstrap estimate, $\sigma_{\rm bs}$, captures statistical variations in averaged profiles, but it does not account for systematic errors that could arise from numerics, from our profile splitting algorithm, from baryonic effects, and (perhaps most importantly) from substructure. We crudely account for such errors by adding a systematic uncertainty of 5\% in quadrature,
\begin{equation}
\sigma_{\rm i,mean/median} = \sqrt{\sigma_{\rm i,bs}^2 + \sigma_{\rm sys,av}^2} = \sqrt{\sigma_{\rm i,bs}^2 + (0.05\ \rho_{\rm i,data})^2} \,.
\end{equation}
The systematic error does have some effect on on the best-fit profiles and parameters. It essentially represents a trade-off: if its value is small, the fit `trusts' the statistically well-constrained inner profile over the noisier transition region; if its value is large, all bins are weighted more or less equally. Given that the tests in \paperone demonstrated convergence to roughly 5\% accuracy, this value is a reasonable approximation to the systematic uncertainty.

We minimize $\chi^2$ using the `trust region reflective' (trf) algorithm of the least-squares function in \textsc{scipy}. We find this algorithm to converge more reliably than the similar `dogbox' variant. Both methods are faster than a conventional Levenberg-Marquardt solver and allow us to impose the parameter limits listed in Table~\ref{table:par_limits}. All parameters are fit in log space to avoid zero or negative values. Despite these optimizations, any steepest-gradient solver is liable to `fall into' the first local minimum in $\chi^2$ it encounters, which may be far from the global minimum. We avoid false minima with the following procedure. We begin from a fiducial initial guess and run the fitter with a relatively low relative target accuracy of $10^{-3}$. Using the output as a starting point, we vary the initial guess of each parameter to its extremes, that is, fractions of $0.05$ and $0.95$ between its lower and upper bounds in log space. If the resulting fit improves on the previous $\chi^2$, we set the new results as the initial guess. If the total gradient at the solution (`optimality') is larger than $10^{-3}$, we repeat the procedure for more initial guesses of the parameter in question, namely fractions of $[0.2, 0.35, 0.5, 0.65, 0.8]$ of its allowed log interval. Once all parameters have been varied, we accept the initial guess that resulted in the lowest $\chi^2$ of all attempts and re-run the fit with a higher accuracy of $10^{-5}$ to obtain the final result. At modest cost, this algorithm ensures that virtually all fits converge to a reasonable solution and that the results are independent of the initial guess. The values for $\chidof$ when fitting the averaged orbiting term hover around unity, though with a large spread that depends on the sample selection and whether we fit mean or median. When fitting the total profiles, $\chidof$ roughly ranges between $0.01$ and $1$.

Fitting individual halo profiles can be particularly challenging because substructure and halo-to-halo scatter can lead to significant deviations from any fitting function \citep[e.g.,][]{umetsu_17}. We thus slightly adjust the procedure laid out in the previous section for individual halo fits, as well as the parameter ranges (Table~\ref{table:par_limits}). Since the simulations do not provide a meaningful estimate of the uncertainty on individual profiles, we assume a Poisson-like statistical uncertainty due to the number of particles as a minimum, 
\begin{equation}
\label{eq:syserr}
\sigma_{\rm i,ind} = \rho_{\rm i,data} \sqrt{\frac{1}{N_{\rm i,ptl}} + \sigma_{\rm sys,ind}^2}  \,,
\end{equation}
where $N_{\rm i,ptl}$ is the number of particles in the given bin and $\sigma_{\rm sys,ind}$ is a fixed systematic error. Since most bins contain $N_{\rm i,ptl} \gg 1$ particles, the normalization of $\chi^2$ is largely determined by the somewhat arbitrary value for $\sigma_{\rm sys,ind}$. This systematic error term effectively weighs bins with low particle number against well-resolved bins because all bins with $N_{\rm i,ptl} \gg 1 / \sigma_{\rm sys,ind}^2$ receive roughly the same weight in the fit. This method prevents the fit from being be dominated by a few bins with high $N_{\rm ptl}$. We set $\sigma_{\rm sys,ind} = 0.25$, meaning that the systematic error dominates for bins with $N_{\rm i,ptl} > 16$. With this value, the median $\chidof$ ranges from $0.5$ to $2$ for the different halo samples. We have verified that the choice of systematic error does not greatly influence the overall distribution of the best-fit parameters.

\subsection{Parameter Limits}
\label{sec:app:fits:limits}

\begin{figure*}
\centering
\includegraphics[trim =  2mm 9mm 0mm 3mm, clip, width=\textwidth]{\figdir/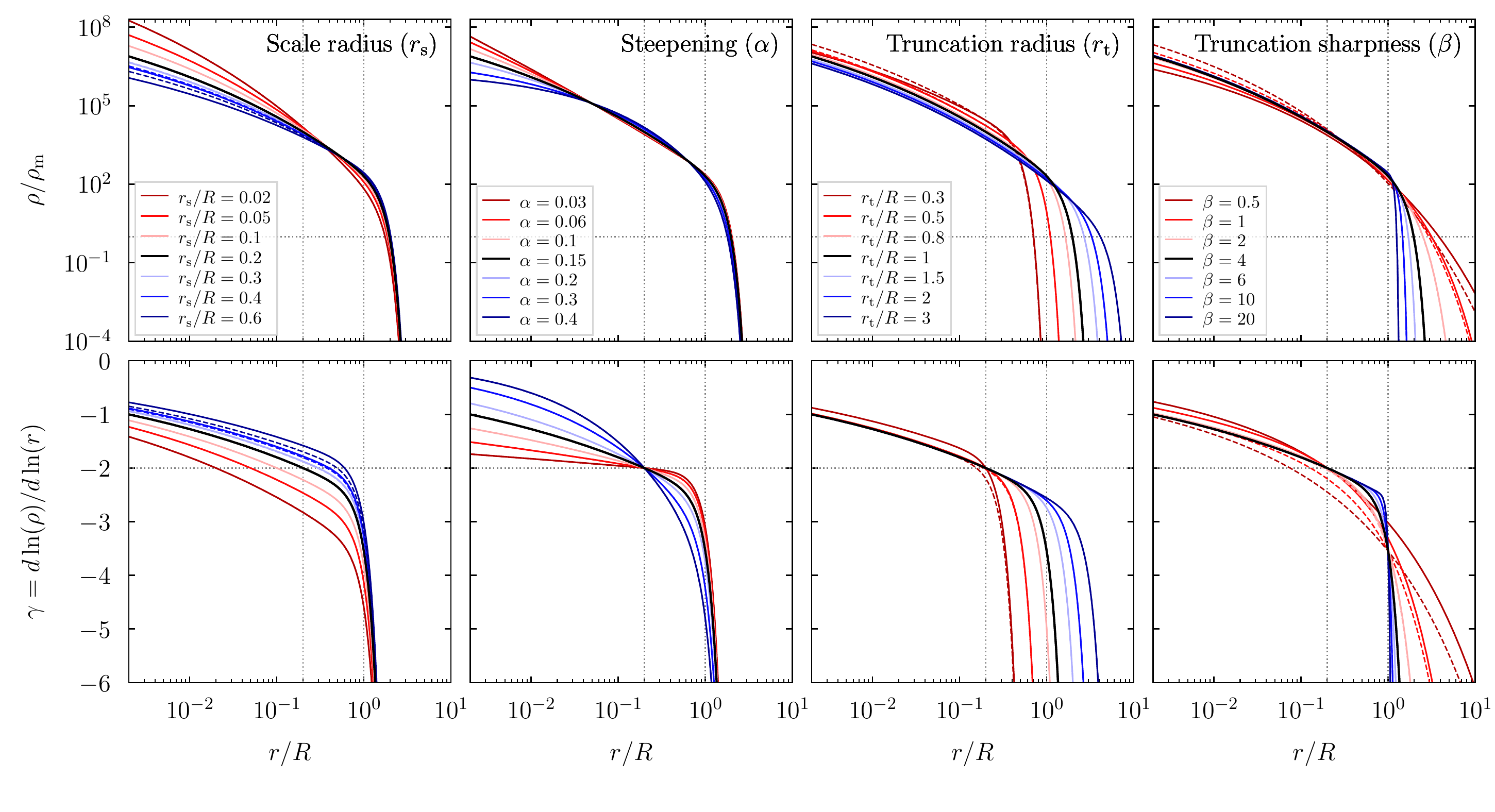}
\caption{Same as Fig.~\ref{fig:moda_pars}, but showing the modified Model B with $\eta = 0.1$ (solid lines) and a comparison to our fiducial Model A (\eqmn{eq:models:moda:s2}, dashed lines). Wherever the dashed lines are invisible, the models make indistinguishable predictions. The most important differences occur in the right two columns. While Model B fixes the slope at $\gamma(\rs) = -2$ regardless of the values of $\rt$ and $\beta$, Model A can exhibit slight deviations from this value if $\rt$ and/or $\beta$ are small.}
\label{fig:modb_pars}
\end{figure*}

The normalization, $\rhos$, is tightly related to the formation redshift of halos \citep{navarro_97, ludlow_13} and thus cannot vary arbitrarily. Nonetheless, we allow a generous range because $\rhos$ is well-constrained in all fits. If we instead use the density at the centre, $\rho_0$, by setting $C = 0$ in Section~\ref{sec:models_orb:moda}, some parameter combinations can extrapolate to extreme central densities and the limits need to be much more flexible (Table~\ref{table:par_limits}).

Some profiles steepen so slowly that they exhibit no true scale radius, i.e., $\gamma \approx -2$ over a wide range of radii. In such cases, the fit tends to shift $\rs$ towards large values that can even exceed $\rt$, which is technically a valid solution because of the symmetry between the $\rs$ and $\rt$ terms in \eqmn{eq:models:moda:s2}. After extensive experimentation, we find it necessary to enforce the correct, physical ordering of the terms by explicitly requiring $\rt > \rs$. In particular, we set $0.01 < \rs / \rtom < 0.45$ and $0.5 < \rt / \rtom < 3$. The range of $\rs$ limits the concentration to $2.2 < \ctom < 100$, but halos outside of this range would correspond to extreme outliers in the population \citep[e.g.,][]{diemer_15}; we never observe a preference for $\rs \gsim 0.4\ \rtom$ in well-constrained fits. The lower limit corresponds roughly to the lowest radii where our averaged profiles are resolved, meaning that $\rs$ cannot be measured reliably near $0.01\ \rtom$. This limit could be revisited with high-resolution simulations that probe smaller radii. Similarly, we do not observe a truncation at radii smaller than $0.5\ \rtom$ in any profiles, and we generally find little evidence for $\rt > 2\ \rtom$. If $\rt$ pushes against the upper limit, the profile does not experience a clear truncation, meaning that the values of $\rt$ and $\beta$ are arbitrary. We have experimented with falling back on an Einasto fit in such cases, but the fit quality is almost always degraded, which demonstrates that almost all orbiting profiles experience some steepening beyond that of the Einasto slope.

For the steepening parameter, we allow $0.03 < \alpha < 0.4$, the most extreme values that could reasonably be ascribed to any profiles we have tested. The lower limit corresponds to an almost invariant slope near $-2$, which is the case for some halos with high accretion rates (\paperone). The Einasto profile can suffer from a well-known degeneracy between $\rs$ and $\alpha$ if profiles do not extend to small radii or if their slope evolves slowly with radius \citep[e.g.,][]{ricotti_07, udrescu_19}. While virtually all averaged profiles do constrain $\alpha$ to some extent, most individual halo profiles do not. In particular, the average fit quality remains roughly constant within the range $0.1 < \alpha < 0.2$, highlighting that individual profiles simply do not contain sufficient information to determine $\alpha$. We could set it based on an $\nu$-$\alpha$ relation \citep{gao_08, klypin_16}, but these trends with peak height turn out to be an artefact of the Einasto form poorly fitting the true, truncated profiles (\paperthree). Instead, we find that $\alpha \approx 0.18$ in most averaged profiles, and that it depends on the accretion rate (\paperthree) and on the slope of the linear power spectrum \citep[\paperone;][]{ludlow_17, brown_20}. While the profiles of individual halos with very high $\Gamma$ do indeed show a mild preference for low $\alpha$, the differences in fit quality are modest at best. Thus, we simply fix $\alpha = 0.18$ in all fits to individual profiles. This value leads to concentrations that are similar to those measured using NFW profiles \citep{dutton_14, ludlow_16}, but other choices have been made in the literature \citep[e.g., $\alpha = 0.16$,][]{wang_20_zoom}. 

We have no intuitive prior for the transition sharpness $\beta$, but the profiles shown in \paperone range from gradual to very sharp transitions. Thus, we allow $\beta$ to fluctuate between $0.1$ and $10$. While some fits technically prefer even larger values, those correspond to an essentially instantaneous cut-off, where $\beta$ is not well defined. We might worry that the $\rt$-$\beta$ term suffers from a degeneracy similar to that of the $\rs$-$\alpha$ term, but we find that $\rt$ and $\beta$ are not intrinsically degenerate and both well-constrained for most averaged profiles (\paperthree). In individual halos, however, the orbiting profiles commonly become unresolved at the radii that most constrain $\beta$. We thus fix $\beta = 3$, which barely decreases the average fit quality of the total profiles. To accommodate profiles that do not exhibit any noticeable truncation, we allow the truncation radius to shift to very large values where it does not matter, $\rt / \rtom \leq 10$.

In the infalling profile, we limit the normalization to $1 < \delone < 100$. Added to the mean density of the Universe, the lower limit corresponds to $\rho_{\rm inf}(\rtom) = 2 \rhom$. Such low overdensities are observed in the self-similar simulation with $n = -1$. We leave the slope to fluctuate between $0.01 < s < 4$, which includes any reasonable (positive) value encountered in simulated profiles. The asymptotic central overdensity $\delmax$ reaches values between $10$ and $1000$ in the profiles where it is constrained, but we allow $1 < \delmax < 10^4$ to allow for profiles that do not exhibit a measurable maximum. We apply the limits discussed so far in the separate fits to the orbiting and infalling profile components. When we combine them to fit the total profiles, we do not vary $\delmax$ because the infalling profile is subdominant to the orbiting term at $r < \rtom$, meaning that $\delmax$ is unconstrained. In a situation where only the total profile is known, $\delmax$ can be set to a reasonable value (e.g., $500$) without a significant impact on the fit.

\section{Model variant with correction at scale radius}
\label{sec:app:modelb}

In Section~\ref{sec:models_orb:moda}, we noted that the truncation term in \eqmn{eq:models:moda:s2}has the undesirable effect of breaking the $\gamma(\rs) = -2$ condition. We now construct a model variant that maintains this condition, which we call Model B (in contrast to the fiducial model, which we shall call Model A for brevity). We wish to introduce a new term in the slope that enforces $\gamma(\rs) = -2$, but we need to be careful. For example, if we add $(\rs/\rt)^\beta$ into the slope of Model A, $\gamma(r)$ is still integrable and adds the term $(\rs / \rt)^\beta \ln(r)$ to $S(r)$ from \eqmn{eq:models:moda:s}. While this new model would now satisfy $\gamma(\rs) = -2$, $\gamma$ has been renormalized at all radii with a fixed term that survives all the way to the centre of the halo, $\gamma(0) = (\rs / \rt)^\beta$. This feature is highly undesirable because it ties the central slope to the entirely unrelated properties of the truncation term. Instead, we introduce a multiplicative term that vanishes at $r = 0$ and approaches $(\rs/\rt)^\beta$ at $r = \rs$,
\begin{equation}
\gamma(r) = -2 \rrsa - \rrtb + \rsrtb \rrse \,.
\end{equation}
We integrate $\gamma / r$ to find
\begin{equation}
S(r) = -\frac{2}{\alpha} \rrsa -\frac{1}{\beta} \rrtb + \frac{1}{\eta} \rsrtb \rrse + C \,.
\end{equation}
Once again, we can set $C = 0$ or ensure $S(\rs) = 0$ by setting
\begin{equation}
\label{eq:models:modb:c}
C = \frac{2}{\alpha} + \left[ \frac{1}{\beta} - \frac{1}{\eta} \right] \rsrtb \,,
\end{equation}
which causes the familiar extra terms from equations~(\ref{eq:models:einasto:s}) and (\ref{eq:models:moda:s}),
\begin{equation}
S = -\frac{2}{\alpha} \left[ \rrsa - 1 \right] -\frac{1}{\beta} \left[ \rrtb - \rsrtb \right] + \frac{1}{\eta} \rsrtb \left[ \rrse - 1 \right] \,.
\end{equation}
While this form looks somewhat complicated, it has a straightforward interpretation: its slope is renormalized to match $-2$ at the scale radius but this normalization factor decays towards $r = 0$. Here, $\eta$ is a nuisance parameter that determines how fast this decay happens: the larger $\eta$, the faster the slope approaches zero at small radii. On the other hand, $\eta$ also introduces an undesirable change in the slope at larger radii (where $r / \rs \gsim \rs / \rt$). However, as long as $\beta \gg 1$ and $\eta < 1$, this correction is barely noticeable. We fix $\eta = 0.1$, but the Model B fits are insensitive in the range $0.01 < \eta < 0.2$. In some fits to mean orbiting-only profiles, the best-fit value of $\alpha$ depends slightly on $\eta$, but this dependence disappears when fitting the entire profile or median orbiting profiles.

Fig.~\ref{fig:modb_pars} shows the impact of different parameter values on Model B and highlights differences to Model A as dashed lines. As expected, Model B produces profiles that are indistinguishable from Model A for all but the most extreme parameter values. Differences are apparent for small $\rt$ and/or small $\beta$. The example of $\rt / \rtom = 0.3$ is outside of the allowed parameter range for our fits ($\rt \geq 0.5 \rtom$), but we do allow very small values of $\beta$. Model A fixes the slope at $\rt$ to be $\gamma(\rt) = -2(\rt/\rs)^\alpha - 1$, which lies between $-3$ and $-4$ for most parameter values. Conversely, Model B fixes the slope at $\rs$ to be $\gamma = -2$ regardless of the other parameters, but the slope at $\rt$ becomes $\gamma(\rt) = -2 (\rt / \rs)^\alpha -1 + (\rt / \rs)^{\eta - \beta}$. The $\beta$-dependent correction term is negligible except for the smallest $\beta$. 

The advantage of Model B is that $\rs$ takes on its usual meaning as the radius where $\gamma = -2$, which avoids parameter degeneracies in power law-like profiles where the scale radius is poorly constrained (\paperone). We have, however, verified that models A and B are indistinguishable for virtually all mean and median profile fits. Wherever there are small differences, Model B tends to fit slightly better. The most significant differences in the best-fit parameters occur for $\rs$ in low-$\Gamma$ averaged samples and in some individual halo profiles. The values of $\rt$ and $\beta$ are almost never affected, except in ill-defined fits. In summary, Model B is, fundamentally, the same profile model as Model A, and the best-fit parameters do not depend on the model except in some pathological cases. While we have exclusively shown Model A fits throughout this paper, we will also investigate best-fit parameters from Model B in \paperthree.

\section{Derivatives with respect to parameters}
\label{sec:app:derivs}

Least-squares fits are often faster and more reliable if the user provides derivatives of the fitted function with respect to its free parameters $\theta$. For orbiting profiles that follow the form $\rho = \rhos e^{S(r)}$, those derivatives become 
\begin{equation}
\dnorm{\ln \rho}{\ln \theta} = \theta \dnorm{S}{\theta} \,.
\end{equation}
The derivative with respect to the normalisation is $\dnorminl{\ln \rho}{\ln \rhos} = 1$ regardless of $S(r)$, and we thus do not repeat it for the models listed below. We have double-checked the following expressions numerically and against Wolfram Alpha.

\subsection{Orbiting: Einasto}
\label{sec:app:derivs:einasto}

If we set $C = 0$, the derivatives with respect to the parameters are
\begin{align}
\label{eq:deriv:einasto:c0}
\dnorm{\ln \rho}{\ln \rs} & = 2 \rrsa \nonumber \\
\dnorm{\ln \rho}{\ln \alpha} & = \frac{2}{\alpha} \rrsa \left[1 - \alpha \ln \rrs \right] \,.
\end{align}
If we set $C = 2 / \alpha$ as in \eqmn{eq:models:einasto:s}, the derivative with respect to $\rs$ remains the same but
\begin{align}
\dnorm{\ln \rho}{\ln \alpha} & = \frac{2}{\alpha} \left[ \rrsa \left[1 - \alpha \ln \rrs \right] - 1 \right] \,.
\end{align}

\subsection{Orbiting: New model (Model A)}
\label{sec:app:derivs:moda}

If $C = 0$, the derivatives with respect to $\rs$ and $\alpha$ are the same as for the Einasto profile with $C = 0$ (\eqmnb{eq:deriv:einasto:c0}), and, analogously,
\begin{align}
\dnorm{\ln \rho}{\ln \rt} & = \rrtb \nonumber \\
\dnorm{\ln \rho}{\ln \beta} & = \frac{1}{\beta} \rrtb \left[ 1 - \beta \ln \rrt \right] \,.
\end{align}
If $C$ is set according to \eqmn{eq:models:moda:c}, we get
\begin{align}
\dnorm{\ln \rho}{\ln \rs} & = 2 \rrsa + \rsrtb \nonumber \\
\dnorm{\ln \rho}{\ln \alpha} & = \frac{2}{\alpha} \left[ \rrsa \left[1 - \alpha \ln \rrs \right] - 1 \right] \nonumber \\
\dnorm{\ln \rho}{\ln \rt} & = \rrtb - \rsrtb \nonumber \\
\dnorm{\ln \rho}{\ln \beta} & = \rrtb \left[ \frac{1}{\beta} - \ln \rrt \right] - \rsrtb \left[ \frac{1}{\beta} - \ln \rsrt \right] \,.
\end{align}

\subsection{Orbiting: New model (Model B)}
\label{sec:app:derivs:modb}

When setting $C = 0$, the parameter derivatives for Model B are
\begin{align}
\dnorm{\ln \rho}{\ln \rs} & = 2 \rrsa + \rsrtb \rrse \left[ \frac{\beta}{\eta} - 1 \right] \nonumber \\
\dnorm{\ln \rho}{\ln \alpha} & = \frac{2}{\alpha} \rrsa \left[ 1 - \alpha \ln \rrs \right]  \nonumber \\
\dnorm{\ln \rho}{\ln \rt} & = \rrtb - \frac{\beta}{\eta} \rsrtb \rrse \nonumber \\
\dnorm{\ln \rho}{\ln \beta} & = \rrtb \left[ \frac{1}{\beta} - \ln \rrt \right] + \frac{\beta}{\eta} \rsrtb \rrse \ln \rsrt \nonumber \\
\dnorm{\ln \rho}{\ln \eta} & = \rsrtb \rrse \left[ \ln \rrs - \frac{1}{\eta} \right] \,.
\end{align}
When setting $C$ as in \eqmn{eq:models:modb:c}, we get
\begin{align}
\dnorm{\ln \rho}{\ln \rs} & = 2 \rrsa + \left( \frac{\beta}{\eta} - 1 \right) \rsrtb \left[ \rrse - 1 \right] \nonumber \\
\dnorm{\ln \rho}{\ln \alpha} & = \frac{2}{\alpha} \left[ \rrsa \left[ 1 - \alpha \ln \rrs \right] - 1 \right] \nonumber \\
\dnorm{\ln \rho}{\ln \rt} & = \rrtb - \rsrtb \left[ \frac{\beta}{\eta} \left[ \rrse - 1 \right] + 1 \right] \nonumber \\
\dnorm{\ln \rho}{\ln \beta} & = \rrtb \left[ \frac{1}{\beta} - \ln \rrt \right] \nonumber \\ 
& - \rsrtb \left[\frac{1}{\beta} - \ln \rsrt \left[ \frac{\beta}{\eta} \left[ \rrse - 1 \right] + 1 \right] \right] \nonumber \\
\dnorm{\ln \rho}{\ln \eta} & = \frac{1}{\eta} \rsrtb \left[ \rrse \left[ \eta \ln \rrs - 1 \right] + 1 \right]\,.
\end{align}

\subsection{Infalling: New model}
\label{sec:app:derivs:plmk}

We write the derivatives in terms of $Q(r)$ as defined in \eqmn{eq:models:plmk:q},
\begin{align}
\label{eq:models:plmax2_derivs}
\dnorm{\ln \rho}{\ln \delone} & = f_{\rm m} \left[ 1 - \frac{1}{Q(r)} \left( \frac{\delone}{\delmax} \right)^{1 / \zeta} \right] \nonumber \\
\dnorm{\ln \rho}{\ln s} & = -f_{\rm m} \frac{s}{Q(r)}  \left( \frac{r}{R} \right)^{s / \zeta} \ln \rrtom \nonumber \\
\dnorm{\ln \rho}{\ln \delmax} & = f_{\rm m} \frac{1}{Q(r)} \left( \frac{\delone}{\delmax} \right)^{1/\zeta} \nonumber \\
\dnorm{\ln \rho}{\ln \zeta} & = f_{\rm m} \frac{\ln(Q)}{\zeta} \left[ \left( \frac{\delone}{\delmax} \right)^{1 / \zeta} \ln \left( \frac{\delone}{\delmax} \right) + s \left( \frac{r}{R} \right)^{s / \zeta} \ln \left( \frac{r}{R} \right) \right]  \,.
\end{align}
All derivatives share the same $f_{\rm m} \equiv 1 - \rhom / \rho(r)$ prefactor, which ensures that they vanish at large radii where $\rho$ asymptotically approaches the mean density. This factor is unity when taking the logarithmic derivative of $\rho - \rhom$, i.e., of only the excess density over the mean. The parameter derivatives of the model without transition sharpness (Section~\ref{sec:models:inf}) are the same but with $\zeta = 1$.

\section{Dynamical properties}
\label{sec:app:comp_df}

In Section~\ref{sec:comp:dynamical}, we investigated the space of binding energy, whose structure is determined by the positive gravitational potential of a spherical mass distribution. We find the potential by numerically integrating
\begin{equation}
\psi(r) = 4 \pi G \left[ \frac{M(r)}{r} + \int_r^\infty \rho(r') r' \rmd r' \right] \,.
\end{equation}
For the Einasto profile, we can numerically reproduce the analytical expression for $\psi$ of \citet{retanamontenegro_12}. Assuming isotropy, the velocity dispersion is found by integrating the Jeans equation,
\begin{equation}
\sigma^2(r) = \frac{1}{\rho(r)} \int_r^\infty \frac{\rho(r') M(r')}{r'^2} \rmd r' \,.
\end{equation}
Analytical expressions for $\sigma$ exist for some profiles, but they are generally complicated and tend not to extend to Einasto (and other non-power law) forms \citep[e.g.,][]{lokas_01}. The distribution function can be found with the expression of \citet{eddington_16}, but this formula involves derivatives of the form $\dnorminl{\rho}{\psi}$, which are not easy to compute for arbitrary systems. We follow an easier route by converting to derivatives in $r$ \citep{binney_82_phasespace, baes_21},
\begin{equation}
f(\varepsilon) = \frac{1}{\sqrt{8} \pi^2} \int_{r_\varepsilon}^\infty \frac{g(r)}{\sqrt{\varepsilon - \psi(r)}} \rmd r \,,
\end{equation}
where $r_\varepsilon$ is the radius where $\psi(r_\varepsilon) = \varepsilon$ and
\begin{equation}
g(r) \equiv \frac{r^2}{M(r)} \left[ \dnormtwo{\rho}{r} + \dnorm{\rho}{r} \left( \frac{2}{r} - \frac{4 \pi \rho(r) r^2}{M(r)} \right) \right] \,.
\end{equation}
The problem of finding $f(\varepsilon)$ has now been reduced to computing $\dnorminl{\rho}{r}$ and $\dnorminltwo{\rho}{r}$. We can generally write
\begin{equation}
\dnorm{\rho}{r} = Q(r) \rho \qquad \mathrm{and} \qquad \dnormtwo{\rho}{r} = \rho \left( \dnorm{Q}{r} + Q^2 \right) \,.
\end{equation}
For the Einasto profile, we compute 
\begin{equation}
Q = -\frac{2}{r} \rrsa \qquad \mathrm{and} \qquad \dnorm{Q}{r} = \frac{Q (\alpha - 1)}{r} \,.
\end{equation}
For Model A, the expressions become somewhat more complicated,
\begin{equation}
Q = -\frac{1}{r} \left[ 2 \rrsa + \rrtb \right]
\end{equation}
and 
\begin{equation}
\dnorm{Q}{r} = -\frac{1}{r} \left( Q + \frac{1}{r} \left[ 2 \alpha \rrsa + \rrtb \right] \right) \,.
\end{equation}
Finding the equivalent expressions for Model B or for the \dkft fitting function would be tedious but straightforward; we omit them because their properties would be almost indistinguishable from Model A. For all profiles, we have checked that we can recover the input density profile from the distribution function via the integral over energy,
\begin{equation}
\rho(r) = 4 \pi \int_0^{\psi(r)} f(\varepsilon) \sqrt{2[\psi(r) - \varepsilon]}\ \rmd \varepsilon \,.
\end{equation}


\bsp
\label{lastpage}
\end{document}